\documentclass[a4paper,fleqn,usenatbib]{mnras}
\usepackage[T1]{fontenc}
\usepackage{ae,aecompl}

\usepackage{graphicx}	
\usepackage{amsmath}	
\usepackage{amssymb}	
\usepackage{booktabs}    
\usepackage{placeins}     
\usepackage{longtable}    

\newcommand{\kms}{km\,s$^{-1}$}
\renewcommand{\ion}[2]{#1$\,${\small \MakeUppercase{\romannumeral #2}}}

\title[Berkeley Sample of Stripped-Envelope SNe]{The Berkeley Sample of Stripped-Envelope Supernovae}

\author[Shivvers et al.]{Isaac Shivvers,$^{1}$\thanks{E-mail: ishivvers@gmail.com}
Alexei V.\ Filippenko,$^{1,2}$
Jeffrey M.\ Silverman,$^{3}$
WeiKang Zheng,$^{1}$  \newauthor
Ryan J.\ Foley,$^{4}$
Ryan Chornock,$^{5}$
Aaron J.\ Barth,$^{6}$
S.\ Bradley Cenko,$^{7,8}$  \newauthor
Kelsey I.\ Clubb,$^{1}$
Ori D.\ Fox,$^{9}$
Mohan Ganeshalingam,$^{10}$
Melissa L.\ Graham,$^{11}$  \newauthor
Patrick L.\ Kelly,$^{1,12}$
Io K.\ W.\ Kleiser,$^{13}$
Douglas C.\ Leonard,$^{14}$
Weidong Li,$^{1,15}$  \newauthor
Thomas Matheson,$^{16}$
Jon C.\ Mauerhan,$^{1,17}$
Maryam Modjaz,$^{18}$  \newauthor
Franklin J.\ D.\ Serduke,$^{1}$
Joseph C.\ Shields,$^{5}$
Thea N.\ Steele,$^{1}$
Brandon J.\ Swift,$^{19,20}$  \newauthor
Diane S.\ Wong,$^{1}$
Heechan Yuk$^{1,21}$
\\ \\
$^{1}$Department of Astronomy, University of California, Berkeley, CA 94720-3411, USA \\
$^{2}$Miller Senior Fellow, Miller Institute for Basic Research in Science, University of California, Berkeley, CA 94720, USA \\
$^{3}$Samba TV, San Francisco, CA 94107, USA \\
$^{4}$Department of Astronomy and Astrophysics, University of California, Santa Cruz, CA 95064, USA \\
$^{5}$Astrophysical Institute, Department of Physics and Astronomy, Ohio University, Athens, OH 45701, USA \\
$^{6}$Department of Physics and Astronomy, 4129 Frederick Reines Hall, University of California, Irvine, CA, 92697-4575, USA \\
$^{7}$Astrophysics Science Division, NASA Goddard Space Flight Center, Mail Code 661, Greenbelt, MD 20771, USA \\
$^{8}$Joint Space-Science Institute, University of Maryland, College Park, MD 20742, USA \\
$^{9}$Space Telescope Science Institute, 3700 San Martin Dr., Baltimore, MD 21218, USA \\
$^{10}$Energy Analysis and Environmental Impacts Division, Lawrence Berkeley National Laboratory, 1 Cyclotron Road, Berkeley, CA 94720, USA \\
$^{11}$Department of Astronomy, University of Washington, 3910 15th Avenue NE, Seattle, WA, 98195, USA \\
$^{12}$Minnesota Institute for Astrophysics, University of Minnesota, 115 Union St. SE, Minneapolis, MN, 55455, USA \\
$^{13}$Cahill Center for Astrophysics, California Institute of Technology, Pasadena, CA 91125, USA \\
$^{14}$Department of Astronomy, San Diego State University, San Diego, CA 92182-1221, USA \\
$^{15}$Deceased 12 December 2011 \\
$^{16}$National Optical Astronomy Observatory, 950 N. Cherry Ave., Tucson, AZ 85719, USA \\
$^{17}$The Aerospace Corporation, 2310 E. El Segundo Blvd., El Segundo, CA 90245, USA \\
$^{18}$Center for Cosmology and Particle Physics, New York University, 4 Washington Place, New York, NY 10003, USA \\
$^{19}$Steward Observatory, University of Arizona, 933 North Cherry Avenue, Tucson, AZ 85721-0065, USA \\
$^{20}$FreeFall Aerospace, 2555 N. Coyote Drive, Tucson, AZ 85745, USA \\
$^{21}$Department of Physics and Astronomy, San Francisco State University, San Francisco, CA 94132, USA \\
}

\date{Accepted for publication in MNRAS}
\pubyear{2018}

\begin{document}
\label{firstpage}
\pagerange{\pageref{firstpage}--\pageref{lastpage}}
\maketitle


\begin{abstract}
We present the complete sample of stripped-envelope supernova (SN) spectra observed by the
Lick Observatory Supernova Search (LOSS) collaboration over the last three decades:
888 spectra of 302 SNe, 652 published here
for the first time, with 384 spectra (of 92 SNe) having photometrically-determined phases.
After correcting for redshift and Milky Way dust reddening and reevaluating the
spectroscopic classifications for each SN,
we construct mean spectra of the three major spectral subtypes (Types IIb, Ib, and Ic)
binned by phase. We compare measures of line strengths and widths made from this sample
to the results of previous efforts, confirming that \ion{O}{1}\,$\lambda$7774
absorption is stronger and found at higher velocity in Type Ic SNe than in Types Ib or IIb SNe
in the first $\sim30$ days after peak brightness, though the widths of nebular emission
lines are consistent across subtypes. We also highlight newly available observations
for a few rare subpopulations of interest.

\end{abstract}


\section{Introduction}

Hydrogen-poor supernovae (SNe) showing strong \ion{Si}{2}\,$\lambda$6355
absorption and presumably
arising from cataclysmic thermonuclear white dwarf explosions
(Type Ia SNe) have long been of intense interest to observers for their cosmological utility
\citep[e.g.,][]{1998AJ....116.1009R,1999ApJ...517..565P},
while a few dozen relatively nearby hydrogen-rich SNe arising from the deaths of massive stars
(Type II SNe) have provided observers a powerful perspective onto the physics of these systems
\citep[e.g.,][]{1993ARA&A..31..175M,1997Natur.386..438M,2016ARA&A..54...19M,2009MNRAS.395.1409S}.
In contrast, hydrogen-poor SNe arising from massive stellar deaths
\citep[the stripped-envelope SNe: Type IIb, with H easily visible at early times and He lines later; Type Ib, with no H but showing obvious He lines; and Type Ic, with neither obvious H nor obvious He --- for more thorough discussions of SN spectral types see][]{1997ARA&A..35..309F,2017hsn..book..195G}
have historically proven less amenable to detailed observational studies than their
hydrogen-rich counterparts, and have been less intensely studied than the cosmologically-useful SNe~Ia.

These stripped-envelope SNe provide a powerful window into the physics that govern
the rapidly evolving and often chaotic end-stages of massive stellar evolution, but many
puzzles remain about their progenitors and host-galaxy environments.
\citet{2012Sci...337..444S} show that $\sim33$\% of massive O-type stars
in the Milky Way undergo significant envelope stripping via Roche-lobe overflow, in good agreement
with the fraction of core-collapse SNe classified as stripped-envelope in a volume-limited sample
\citep{2011MNRAS.412.1522S,2017PASP..129e4201S}; it appears that the most common progenitors
of normal SNe~IIb, Ib, and Ic arise within binary systems \citep[e.g.,][]{2010ApJ...725..940Y}.
However, stripped-envelope SNe are heterogenous; it is likely that several progenitor channels contribute
to the observed SN populations. For example, a small fraction of the SNe~Ib likely arise from isolated and very massive
progenitors that have lost their envelopes through a vigorous line-driven wind
\citep[e.g., PTF10qrl and iPTF15cna;][]{2018arXiv180700100F,1975MSRSL...9..193C}, while at least some members of the
SN~Ibn subclass arise from systems where explosive outbursts have contributed to their envelope stripping
\citep[e.g., SN~2006jc;][]{2007ApJ...657L.105F}.

Even when restricting the discussion to the relatively common classes (Types IIb, Ib, and Ic), many
questions remain.  The key observational indicators used to discriminate between subtypes exhibit
confounding traits: the strength of the H$\alpha$ feature in SN spectra is a sensitive and easily-saturated
indicator of the presence (or lack) of hydrogen, making it difficult to discern whether SNe~Ib and IIb are
truly separate populations or instead lie along a continuum with the hydrogen-rich Type II SNe,
while the optical features of \ion{He}{1} may not be present though the ejecta are helium-rich
\citep[e.g.,][]{2012MNRAS.424.2139D}.
These effects sometimes make the physical interpretation of observational trends between subtypes
difficult, but careful studies of the spectra of these diverse and multifarious populations help us
to tease apart their complexities.

Many single-object papers examining stripped-envelope SNe have been written, but for years the work of
\citet{2001AJ....121.1648M}, analyzing 84 spectra of 27 SNe, provided researchers
the only large multi-object sample of optical spectra.
The more recent work of \citet{2014AJ....147...99M}, however, has
greatly expanded the set of publicly-available data by publishing 645 spectra of 73 SNe,
and the very recent work of \citet{2018arXiv180700100F} has contributed a valuable analysis of 507 spectra of 173 events,
all having a measured date of peak brightness.
Interest in the study of the bulk properties of these relatively poorly-understood events
has been reignited by these samples
\citep[e.g.,][]{2016ApJ...827...90L,2016ApJ...832..108M, 2018A&A...609A.134S,2018A&A...609A.135S,2018A&A...609A.136T}.

This paper builds on past efforts to understand stripped-envelope SNe by presenting the largest
single dataset yet released (including numerous spectra obtained at late times)
and analysing this dataset to interpret the underlying physical mechanisms
driving the subtype divisions among stripped-envelope SNe.
The observational study of all SNe via their spectra has long been the focus of
our supernova research group at U.C.\ Berkeley.
Over the last three decades we have deployed the resources of the Lick and Keck Observatories toward
this goal, and with this paper we present essentially all stripped-envelope SN spectra observed by the
University of California (U.C.) Berkeley / Lick Observatory Supernova Search \citep[LOSS; ][]{2001ASPC..246..121F}
effort, from 1985  up through the end of 2015,
including 888 spectra of 302 SNe.  A fraction of these data (236 spectra) have already been examined,
either by \citet[][who published the pre-2000 Berkeley data]{2001AJ....121.1648M}
or as part of single-object studies, but we include
them here again to provide a coherent dataset observed with similar observational
techniques and reduced via similar procedures.  A date of peak brightness
can be measured for 92 of the SNe having spectra published here, either from our own
internal photometric database or previously published data, allowing us to label 384
spectra with their phases.  These spectra are made publicly available through the
Berkeley Supernova Database\footnote{\url{heracles.astro.berkeley.edu/sndb/}}
\citep[the SNDB, first described by][]{2012MNRAS.425.1789S},
the Weizmann Interactive Supernova data REPository\footnote{\url{https://wiserep.weizmann.ac.il/}}
\citep[WiseREP,][]{2012PASP..124..668Y},
and the Open Supernova Catalog\footnote{\url{https://sne.space/}} \citep{2017ApJ...835...64G}.
Here we follow similar efforts from the past few years in which have examined and made publicly available
the accumulated Berkeley samples
of Type Ia SNe \citep{2012MNRAS.425.1789S,2012MNRAS.425.1819S,2012MNRAS.425.1889S,2012MNRAS.425.1917S,2013MNRAS.430.1030S}
and Type II SNe \citep{2014MNRAS.442..844F,2014MNRAS.445..554F}.

This paper is laid out as follows.
Section \ref{sec:data} describes our observational methods and data-reduction
procedures, including our efforts to determine robust spectral types and phases.
We describe and present the dataset itself in Section \ref{sec:sample}.
In Section \ref{sec:results}, we compare analyses of this sample to
previously-published works, we present continuum-normalised mean spectra of SNe~IIb, Ib, and Ic in
four different phase bins, and we discuss a few rare subpopulations of interest that are represented
in this dataset.

\section{Observations, Data Reductions, and Calibrations}
\label{sec:data}

Our data include observations obtained with a variety of instruments mounted
on the telescopes at Lick and Keck Observatories:
the UV Schmidt spectrograph \cite[1987--1992;][]{schmidt} and
the Kast Double Spectrograph \citep[since 1992;][]{kast}
on the Shane~3\,m telescope, and the
Low Resolution Imaging Spectrometer \citep[LRIS;][]{1995PASP..107..375O},
the Echellette Spectrograph and Imager \citep[ESI;][]{2002PASP..114..851S},
and the DEep Imaging Multi-Object Spectrograph \citep[DEIMOS;][]{2003SPIE.4841.1657F},
on the two Keck~10\,m telescopes.

Taking advantage of a few nights on the Shane~3\,m every lunar cycle,
augmented by occasional nights with one of the Keck telescopes, our group strives to maintain a steady
cadence of observations for all of the SNe we study --- \citet{2012MNRAS.425.1789S}
describe our observing strategies in more detail.
For the sample of spectra presented here, we calculate a typical (median) time lag of 14.8\,days
between successive observations of any given SN.

With few exceptions, our spectra are flat-field corrected with dome flats and
wavelength calibrated with observations of emission-line calibration lamps
obtained each night (usually at the position of the science target),
flux-calibrated via spectra of standard stars taken at an airmass comparable
to that of the science target, and observed at the parallactic angle
\citep{1982PASP...94..715F} to minimise the effects of atmospheric dispersion.

The data-reduction procedures we use to transform the two-dimensional
spectrogram measured by a CCD into a wavelength-calibrated,
flux-calibrated, one-dimensional spectrum are similar across
instruments and have changed only slowly over time ---
\citet{2012MNRAS.425.1789S} discuss these procedures in detail.
The codes used to perform these reductions consist of a mixture of
{\tt IRAF},\footnote{IRAF is distributed by the National Optical Astronomy Observatory, which is
operated by the Association of Universities for Research in Astronomy,
Inc., under cooperative agreement with the National Science Foundation.}
 {\tt IDL}, and {\tt Python} and are
publicly available.\footnote{\url{https://github.com/ishivvers/TheKastShiv}}

\subsection{Dates of Peak Brightness via Photometry}
\label{sec:peaks}

Though precise determinations of the explosion dates are only rarely obtained
for SNe, the date of peak brightness can often be observationally
constrained. Here we attempt to date all of our spectra relative
to the SN's epoch of photometric peak so as to compare SN evolution across different events.
For the SNe in our sample, we accumulate the light curves and observed dates of photometric peak
we can find in the published literature or in our own (both published and unpublished)
repository of photometry.\footnote{Our photometry was obtained with the 0.76\,m
Katzman Automatic Imaging Telescope \citep[KAIT;][]{2001ASPC..246..121F} at Lick Observatory
and, to a much lesser degree, the Lick 1\,m Nickel telescope.}

\begin{table}
\begin{center}
\caption{Phase of Peak by Passband}
\label{tab:t6000}
\begin{tabular}{ c c c }
\toprule
Passband & $\Delta t$ $\pm 1\sigma$$^\dagger$ & \# SNe \\
 & (days) & \\
\midrule
$U$ & $-2.2 \pm 1.7$ & 6 \\
$B$ & $-2.6 \pm 1.2$ & 17 \\
$V$ & $-0.7 \pm 0.5$ & 20 \\
clear & $0.3 \pm 0.7$ & 20 \\
$R$ & $0.6 \pm 0.6$ & 20 \\
$I$ & $1.6 \pm 1.4$ & 20 \\
$J$ & $6.9 \pm 2.5$ & 7 \\
$H$ & $8.1 \pm 3.6$ & 7 \\
$K$ & $9.1 \pm 4.7$ & 6 \\
\bottomrule
$^\dagger$Relative to $t_{6000}$.
\end{tabular}
\end{center}
\end{table}

\begin{figure}
    \includegraphics[width=0.45\textwidth]{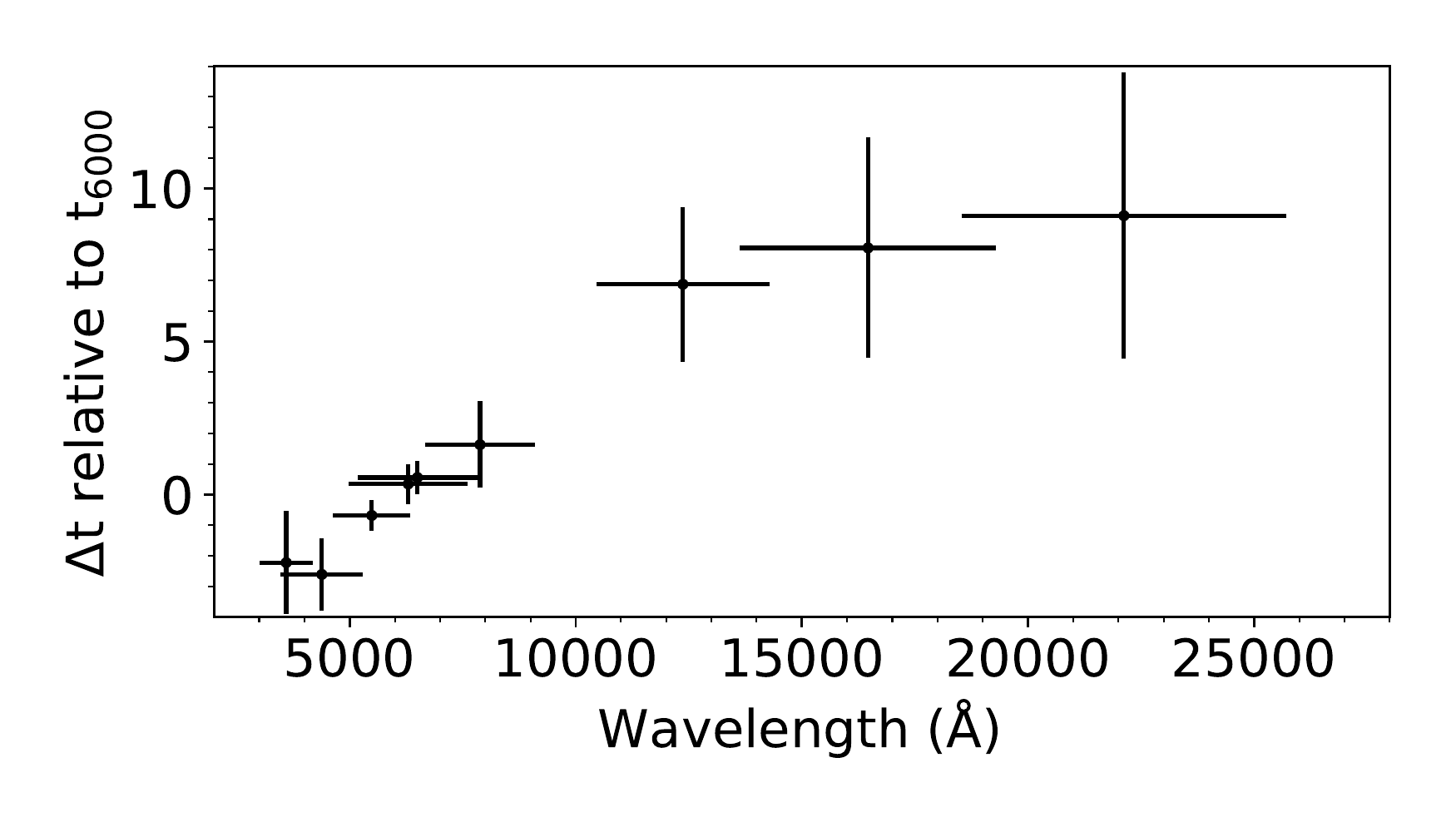}
    \caption{Offsets between peak brightness as measured in each passband
    and the peak at 6000\,\AA\ ($t_{6000}$).  Error bars in $\Delta t$
    represent the standard deviation across measures from all SNe, and
    we illustrate the effective widths of each passband using error bars
    in wavelength.}
    \label{fig:t6000}
\end{figure}

Several studies have documented a wavelength dependency of the measured time
of peak brightness \citep[e.g.,][]{2011ApJ...741...97D,2014ApJS..213...19B},
with bluer passbands peaking before the red, and our accumulated data support those trends.
Our data come from diverse sources, with the dates of peak constrained in different passbands for different events,
and so we fit for this wavelength dependence and strive to place all of these different measures on a similar scale.

First, we collect the published dates of peak brightness from all available literature sources.
If error bars on the value were not published (as was often the case) we attempted
to obtain the light-curve data and fit for the peak ourselves.
Second, we collected all as-yet unpublished photometric data that constrain the date of peak for these SNe from our
{\tt SNDB} repository. To fit for a peak, we require a light curve with at least two data points showing the rise and turnover.
(The declines of these light curves are generally well-sampled, but the early rise and the peak itself are often missed.)
We model the light curve with a  quadratic polynomial and we restrict our fit to observations taken
within $\pm 10$\,days of peak brightness.

We then chose a specific wavelength at which to normalise the diverse light-curve passbands;
examining our dataset, we find that normalising all data to the date of peak brightness as measured at 6000\,\AA\
produces a reasonable compromise by minimising the number of SNe for which we must significantly extrapolate in wavelength
\citep[6000\,\AA\ is close to the central wavelengths of both the KAIT {\it clear} and {\it R} passbands;][]{2003PASP..115..844L}.
To do this, we take all SNe with estimates of peak measured in four or more passbands with central wavelengths
both above and below 6000\,\AA, and we then fit for the wavelength dependence of the peak-brightness date for each SN.
We utilise Gaussian-process regression to propagate our uncertainties through the interpolation
and produce measures and uncertainty estimates for
the date at which the SN light curve peaked in the virtual passband centered at 6000\,\AA, $t_{6000}$.

For each individual light curve, we calculate the time delay between $t_{6000}$ and the date of peak brightness
through the given passband ($\Delta t$).  For each passband, we then take the weighted average of the $\Delta t$
found across different SNe to infer that passband's phase of peak relative to $t_{6000}$,
and we report the standard deviation as an estimate of our uncertainties.
See Table~\ref{tab:t6000} and Figure~\ref{fig:t6000} for our results and the number of SNe used to reach them.

\subsection{Spectroscopic Classifications}
\label{sec:classification}

The exact subtype definitions best used to discriminate between stripped-envelope SNe and
highlight their underlying physics have long been debated and discussed in the literature
\citep[e.g.,][]{1990AJ....100.1575F,1990RPPh...53.1467W,1994ApJ...436L.135W,1996ApJ...462..462C,2001AJ....121.1648M,2006PASP..118..791B},
and updates to previously-announced spectral types are often justified when performing
comparisons between types \citep[e.g.,][]{2014AJ....147...99M,2016arXiv160902922S}.
We therefore strive to confirm and, if required, update the spectral type for all of the SNe in our sample.

We rely heavily upon the SuperNova IDentification code \citep[{\tt SNID};][]{2007ApJ...666.1024B}
and follow the classification guidelines laid out by \citet{2016arXiv160902922S}.  We use the
BSNIP v7.0 {\tt SNID} templates \citep{2012MNRAS.425.1789S} augmented by the \citet{2014arXiv1405.1437L}
stripped-envelope templates (and following all suggestions from their Table 4).
Given the difficulty of classifying historical SNe having limited observational coverage, we assign a single type only when possible
(using all data available on each SN, both ours and previously-published observations);
if multiple types are reasonable we list them all and do not choose between them.

We follow a multi-step process for each of the spectra in our sample, iteratively attempting to obtain spectral
type and subtype classifications and (if possible) age estimates.  For each SN represented in our sample,
we then examine the set of results found for all of the spectra of the object and choose the best classification.
In detail, we undertake a multistep process for each spectrum similar to that described by \citet{2012MNRAS.425.1789S}.
We first run {\tt SNID}, setting the redshift to that of SN's host galaxy (using the {\tt forcez} keyword) and demanding a
minimum {\tt rlap} value of 10.  If this produces a single preferred type (if the type of the best-match template matches
the most common type amongst all spectra which pass our {\tt rlap} threshold, and if $>50$\% of the templates passing
that threshold are of the same type), we record the preferred type. If there are not at least three matches with
{\tt rlap} $ > 10$, we relax the threshold to {\tt rlap} $ > 5$ and repeat.  If the above process does not produce
a single preferred classification we record the set of reasonable types and move on.

However, if a single type is successfully determined, we then attempt to determine a subtype. We again run {\tt SNID}
(demanding {\tt rlap} $ > 10$) and consider only correlations with templates of the type as determined above
(taking advantage of the {\tt usetype} keyword) and again relaxing the {\tt rlap} threshold to 5, if needed.
If a single subtype is preferred (under the same requirements as above), we adopt it; if not, we do not
record a subtype classification for this spectrum.

We then accumulate all of the spectrum-specific classification results for each SN and determine whether a single classification
is preferred for each object. For those SNe with conflicting type classifications amongst their various spectra, we prefer the classifications
obtained at phases between peak brightness and $+$50\,d post-peak (as indicated by either light-curve peak measures or the ages of the best-match
{\tt SNID} templates; see below), since those are the phases that are observationally best constrained and which have the best coverage
in the {\tt SNID} template set.

If spectra exist only at phases where distinctions are difficult to draw, we record multiple possible types.
If multiple conflicting classifications (or no trustworthy classifications) exist for a single object, we examine all available data in more detail,
and if a convincing classification can be obtained by incorporating light-curve information, we do so \citep[in the same manner as][]{2016arXiv160902922S}.
Note that few {\tt SNID} templates exist for many of the rarer subtypes (see \S\ref{sec:rare}). In most cases, the spectra for these
subtypes resulted in conflicting classifications each having low {\tt rlap} values, and the examples discussed in \S\ref{sec:rare}
were predominantly discovered and classified by hand after our automatic classification methods failed.
If multiple types remain plausible, we list them all; the resulting type assignments are given in Table~\ref{tab:sne}, and
for several SNe they differ from those first announced in the International Astronomical Union Circulars (IAUCs; also CBETs) and Astronomer's Telegrams (ATels).

\subsection{Dates of Peak Brightness via {\tt SNID}}
\label{sec:snidpeaks}

For those SNe without a photometrically-measured date of peak brightness, but
with a single clearcut type and subtype classification, we use {\tt SNID}
(and its library of template spectra having photometrically-determined dates of peak brightness)
to estimate the age of the SN at each epoch of spectroscopy.
We rerun {\tt SNID}, comparing only against templates of the correct type and subtype, and demanding
{\tt rlap}$ \ge 0.75 \times ${\tt rlap}$_{\rm best}$, where {\tt rlap}$_{\rm best}$ is the {\tt rlap} value of the best-match template.
For each SN, we accumulate the implied dates of peak brightness based upon the ages estimated for the spectra.
If there are three or more such estimates, we take their median as our best estimate.

\begin{figure}
    \includegraphics[width=0.45\textwidth]{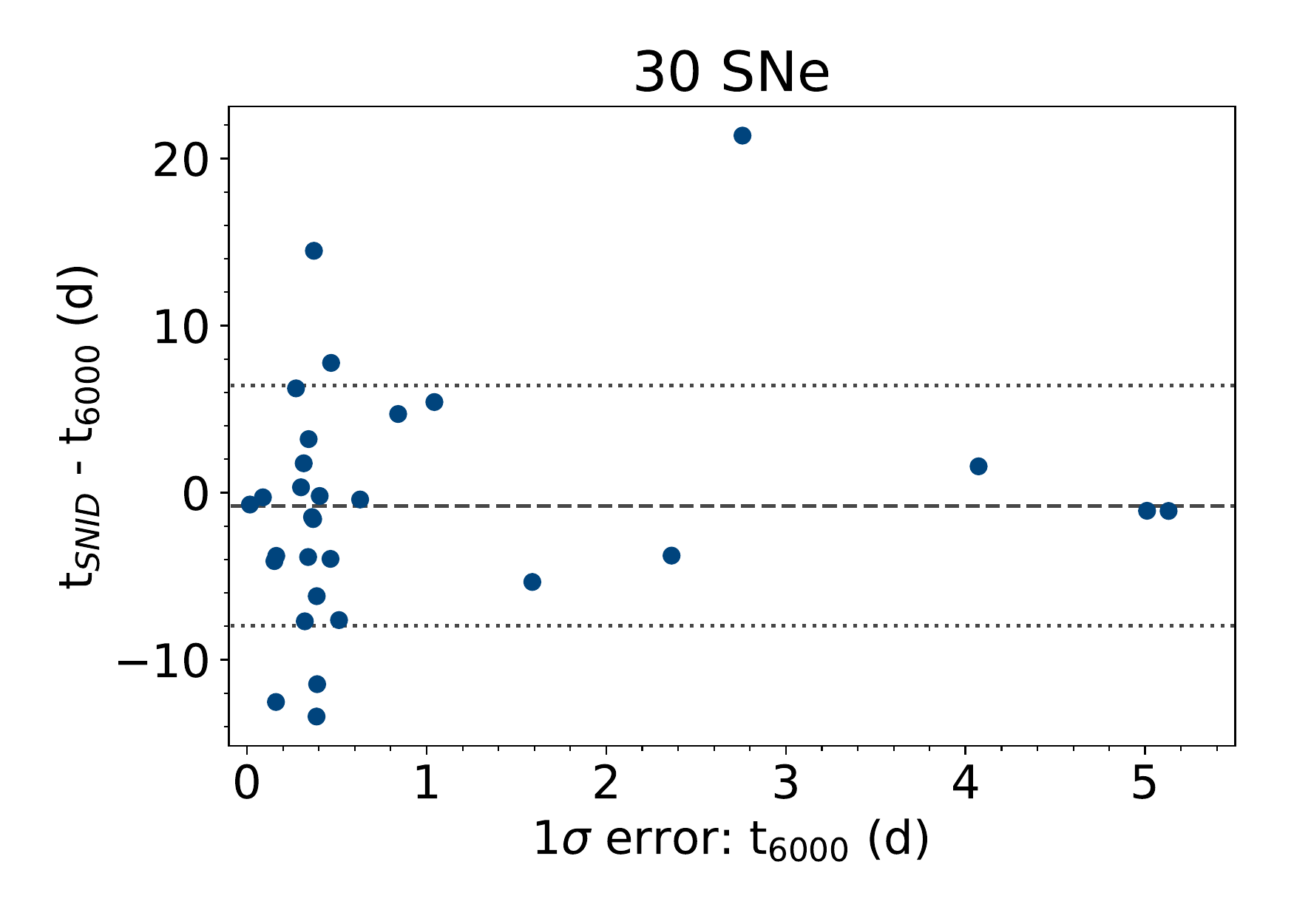}
    \caption{The difference between dates of peak brightness as predicted
    by {\tt SNID} ($t_{\rm SNID}$) and as inferred at 6000\,\AA\ ($t_{6000}$),
    plotted against the 1$\sigma$ error bars on the measured date of peak brightness.}
    \label{fig:peak_lc_vs_snid}
\end{figure}

We assess the quality of our {\tt SNID}-determined peak-brightness dates
with the subset of SNe having both {\tt SNID}-determined and photometrically-measured peak dates.
Amongst the 30 events with both of the above, the mean discrepancy was 0.8\,days
({\tt SNID} has a slight bias toward predicting a date of peak brightness earlier than that measured from the light curve),
and the distribution of differences exhibited a standard deviation of 7.2\,days; see
Figure~\ref{fig:peak_lc_vs_snid}.
For the 31 SNe without a photometrically-measured date of peak brightness and with a {\tt SNID}-determined one,
we adopt the {\tt SNID} value with estimated error bars of $7.2$\,d
(in most of these cases, the SN was first discovered post-peak).

\subsection{Redshift and Dust Reddening Corrections}
\label{sec:redcorrect}

We apply redshift corrections to all of our spectra and analyse them in their host galaxy's
rest frame.  For all but a few SNe in our sample, we are able to associate the event with
its host galaxy and adopt the host's previously-published redshift measures, and for most of the remaining
events we are able to measure the redshift directly from narrow galaxy emission lines within
our spectra.  However, both of these methods fail for SNe 2008gk and 2009ka; for these two events we
infer the redshift via spectral cross-correlation using {\tt SNID}.

We take the line-of-sight extinction arising within our Galaxy
from the dust maps of \citet{2011ApJ...737..103S}, and we assume $R_V=3.1$
to correct each spectrum for Milky Way dust reddening \citep{1989ApJ...345..245C,1994ApJ...422..158O}.
The magnitude and wavelength dependance of dust reddening arising from
within each SN's host galaxy are largely unknown, however.
A small number of spectra show narrow \ion{Na}{1}\,D absorption lines from the host,
but most do not, and doubt exists about the predictive power of this feature in low-resolution spectra
like these \citep{2011MNRAS.415L..81P,2013ApJ...779...38P}.
We therefore do not attempt to correct for reddening
within the host galaxy, and instead endeavor to perform analyses that do not
depend upon exact flux determinations.

\section{The Sample}
\label{sec:sample}

\begin{table*}
\centering
\caption{The SNe in our sample.}
\label{tab:sne}

\begin{tabular}{ l | c c c c c c }
\toprule
Name & Type & Host & Redshift & E(B-V)$_{\rm MW}$ & Date of Peak        & \# Spectra \\
     &      &      &          &                   & (MJD $\pm 1\sigma$) &            \\
\midrule
SN 1985F & Ib & NGC 4618 & 0.00180$^{a}$ & 0.018 & - & 1 \\ 
SN 1987K & IIb & NGC 4651 & 0.00266$^{b}$ & 0.023 & $47009.6 \pm 7.2^{c}$ & 7 \\ 
SN 1987M & Ic & NGC 2715 & 0.00441$^{d}$ & 0.022 & - & 5 \\ 
SN 1988L & Ic & NGC 5480 & 0.00638$^{e}$ & 0.016 & - & 3 \\ 
SN 1990aa & Ic & MCG+05-03-16 & 0.01647$^{d}$ & 0.046 & - & 3 \\ 
SN 1990aj & Ib/Ic & NGC 1640 & 0.00536$^{f}$ & 0.030 & - & 1 \\ 
SN 1990B & Ic & NGC 4568 & 0.00750$^{g}$ & 0.028 & - & 4 \\ 
SN 1990U & Ib & NGC 7479 & 0.00794$^{h}$ & 0.097 & $48094.2 \pm 7.2^{c}$ & 13 \\ 
SN 1991A & Ic & IC 2973 & 0.01059$^{d}$ & 0.017 & $48246.0 \pm 7.2^{c}$ & 6 \\ 
SN 1991ar & Ib & IC 49 & 0.01521$^{h}$ & 0.019 & $48490.6 \pm 7.2^{c}$ & 2 \\ 
SN 1991D & Ib & LEDA 84044 & 0.04175$^{b}$ & 0.053 & - & 2 \\ 
SN 1991L & Ic & MCG+07-34-134 & 0.03054$^{e}$ & 0.012 & - & 1 \\ 
SN 1991N & Ic & NGC 3310 & 0.00300$^{i}$ & 0.019 & - & 4 \\ 
SN 1993J & IIb & NGC 3031 & -0.00047$^{j}$ & 0.069 & $49095.5 \pm 0.6^{k}$ & 50 \\ 
SN 1994ai & Ib/Ic & NGC 908 & 0.00500$^{f}$ & 0.022 & - & 1 \\ 
SN 1994I & Ic & NGC 5194 & 0.00150$^{g}$ & 0.031 & $49451.3 \pm 0.1^{l}$ & 15 \\ 
SN 1995bb & Ib/Ic & Anon J001618+1224 & 0.00580$^{m}$ & 0.096 & - & 1 \\ 
SN 1995F & Ib & NGC 2726 & 0.00487$^{b}$ & 0.031 & $49772.8 \pm 7.2^{c}$ & 4 \\ 
SN 1996cb & IIb & NGC 3510 & 0.00240$^{g}$ & 0.026 & $50452.7 \pm 1.0^{n}$ & 4 \\ 
SN 1997B & Ic & IC 438 & 0.01041$^{h}$ & 0.062 & - & 1 \\ 
SN 1997dc & Ib & NGC 7678 & 0.01167$^{d}$ & 0.042 & - & 1 \\ 
SN 1997dd & IIb & NGC 6060 & 0.01466$^{e}$ & 0.070 & - & 1 \\ 
SN 1997dq & Ic & NGC 3810 & 0.00300$^{o}$ & 0.038 & - & 4 \\ 
SN 1997ef & Ic-BL & UGC 4107 & 0.01169$^{h}$ & 0.036 & - & 6 \\ 
SN 1997ei & Ic & NGC 3963 & 0.01060$^{p}$ & 0.019 & - & 2 \\ 
SN 1997X & Ib & NGC 4691 & 0.00375$^{q}$ & 0.024 & $50483.8 \pm 7.2^{c}$ & 1 \\ 

\multicolumn{7}{c}{{\it (Truncated; full table available online)}} \\

\bottomrule
\multicolumn{7}{|p{0.75\textwidth}|} {{\bf References:} [a]: \citet{2012AnA...538A.120L}, [b]: \citet{2002LEDA.........0P}, [c]: {\tt SNID}, [d]: \citet{1999PASP..111..438F}, [e]: \citet{2011yCat.2306....0A}, [f]: \citet{2004MNRAS.350.1195M}, [g]: \citet{2002PASP..114..833P}, [h]: \citet{2011ApJ...731L...4M}, [i]: \citet{2003AnA...401..927B}, [j]: \citet{2000AJ....120.1499M}, [k]: \citet{1994AJ....107.1022R}, [l]: \citet{1996AJ....111..327R}, [m]: \citet{2012AnA...544A..81H}, [n]: \citet{1999AJ....117..736Q}, [o]: \citet{2006AnA...447..121B}, [p]: \citet{2005ApJS..160..149S}, [q]: \citet{2008yCat.2282....0A}.}
\end{tabular}

\end{table*}

\begin{table*}
\centering
\caption{The spectra presented in this paper.}
\label{tab:data}
\tiny
\begin{tabular}{ l | c c c c c c c }
\toprule
Name & SN Type & Date & Phase$^{\alpha}$ & Instrument & Wavelength Range & Resolution$^{\beta}$ & Reference$^{\gamma}$ \\
     &         & (UT) & (days)           &            & (\AA)            & (\AA)              & \\
\midrule
SN 1985F$^{\delta}$ & Ib & 1985-04-01 & - & Double Spectrograph/ITS & 3180 - 10100 & 12 & \citet{1986AJ.....91..691F} \\ 
SN 1987K & IIb & 1987-07-31 & $-1.8 \pm 7.2$ & UV Schmidt & 4280 - 7140 & 12 & \citet{1988AJ.....96.1941F} \\ 
SN 1987K & IIb & 1987-08-01 & $-0.8 \pm 7.2$ & UV Schmidt & 4280 - 7140 & 12 & \citet{1988AJ.....96.1941F} \\ 
SN 1987K & IIb & 1987-08-07 & $5.2 \pm 7.2$ & UV Schmidt & 3200 - 8880 & 12 & \citet{1988AJ.....96.1941F} \\ 
SN 1987K & IIb & 1987-08-09 & $7.2 \pm 7.2$ & UV Schmidt & 4510 - 7570 & 12 & \citet{1988AJ.....96.1941F} \\ 
SN 1987K & IIb & 1987-08-12 & $10.2 \pm 7.2$ & UV Schmidt & 5960 - 7570 & 12 & \citet{1988AJ.....96.1941F} \\ 
SN 1987K & IIb & 1987-12-25 & $145.2 \pm 7.2$ & UV Schmidt & 6120 - 9270 & 12 & \citet{1988AJ.....96.1941F} \\ 
SN 1987K & IIb & 1988-02-24 & $206.2 \pm 7.2$ & UV Schmidt & 6040 - 9230 & 12 & \citet{1988AJ.....96.1941F} \\ 
SN 1987M & Ic & 1987-09-28 & - & UV Schmidt & 3130 - 10000 & 12 & \citet{1990AJ....100.1575F} \\ 
SN 1987M & Ic & 1987-11-22 & - & UV Schmidt & 3280 - 9100 & 12 & \citet{1990AJ....100.1575F} \\ 
SN 1987M & Ic & 1987-12-26 & - & UV Schmidt & 3180 - 9070 & 12 & \citet{1990AJ....100.1575F} \\ 
SN 1987M & Ic & 1988-02-25 & - & UV Schmidt & 3310 - 9220 & 12 & \citet{1990AJ....100.1575F} \\ 
SN 1988L & Ic & 1988-06-28 & - & UV Schmidt & 3500 - 9300 & 12 & \citet{2001AJ....121.1648M} \\ 
SN 1988L & Ic & 1988-07-17 & - & UV Schmidt & 5970 - 9150 & 12 & \citet{2001AJ....121.1648M} \\ 
SN 1988L & Ic & 1988-09-15 & - & UV Schmidt & 6000 - 9100 & 12 & \citet{2001AJ....121.1648M} \\ 
SN 1990aa & Ic & 1990-09-27 & - & UV Schmidt & 3900 - 9850 & 12 & \citet{2001AJ....121.1648M} \\ 
SN 1990aa & Ic & 1990-10-20 & - & UV Schmidt & 3900 - 7020 & 12 & \citet{2001AJ....121.1648M} \\ 
SN 1990aa & Ic & 1991-01-23 & - & UV Schmidt & 3900 - 9800 & 12 & \citet{2001AJ....121.1648M} \\ 
SN 1990aj & Ib/Ic & 1991-03-10 & - & UV Schmidt & 5800 - 8950 & 12 & \citet{2001AJ....121.1648M} \\ 
SN 1990B & Ic & 1990-03-25 & - & UV Schmidt & 3950 - 9800 & 12 & \citet{2001AJ....121.1648M} \\ 
SN 1990B & Ic & 1990-04-01 & - & UV Schmidt & 3950 - 9800 & 12 & \citet{2001AJ....121.1648M} \\ 
SN 1990B & Ic & 1990-06-15 & - & UV Schmidt & 6720 - 9850 & 12 & \citet{2001AJ....121.1648M} \\ 
SN 1990B & Ic & 1990-04-30 & - & UV Schmidt & 3930 - 9800 & 12 & \citet{2001AJ....121.1648M} \\ 
SN 1990U & Ib & 1990-07-30 & $8.7 \pm 7.2$ & UV Schmidt & 3120 - 7500 & 12 & \citet{2001AJ....121.1648M} \\ 
SN 1990U & Ib & 1990-07-31 & $9.7 \pm 7.2$ & UV Schmidt & 3120 - 9850 & 12 & \citet{2001AJ....121.1648M} \\ 
SN 1990U & Ib & 1990-08-29 & $38.7 \pm 7.2$ & UV Schmidt & 3900 - 9850 & 12 & \citet{2001AJ....121.1648M} \\ 

\multicolumn{8}{c}{{\it (Truncated; full table available online)}} \\
\midrule
\multicolumn{8}{l}{$^{\alpha}$Days relative to peak brightness, if date of peak is constrained.} \\
\multicolumn{8}{l}{$^{\beta}$Estimated average resolution across the spectrum.} \\
\multicolumn{8}{l}{$^{\gamma}$Original publications for already-published spectra.} \\
\multicolumn{8}{l}{$^{\delta}$This historical spectrum was observed under unique conditions ---
                              we exclude from our analyses below but we include it here for completeness.} \\
\bottomrule
\end{tabular}

\end{table*}

Table~\ref{tab:sne} presents the sample of SNe included in this data release, our adopted
classifications and dates of peak brightness (if available),
and the values we used to perform redshift and dust-reddening corrections.
Table~\ref{tab:data} provides a description of the spectra included in our sample.
Our research group utilises a database of observations and associated metadata
\citep[the SNDB;][]{2012MNRAS.425.1789S}
to orchestrate our observational campaigns, to facilitate access to
and analysis of our large repository of data, and to share those
data publicly --- all of the spectra described herein are publicly
available through the SNDB web page.\footnote{\url{heracles.astro.berkeley.edu/sndb/}}

\begin{figure}
    \includegraphics[width=0.45\textwidth]{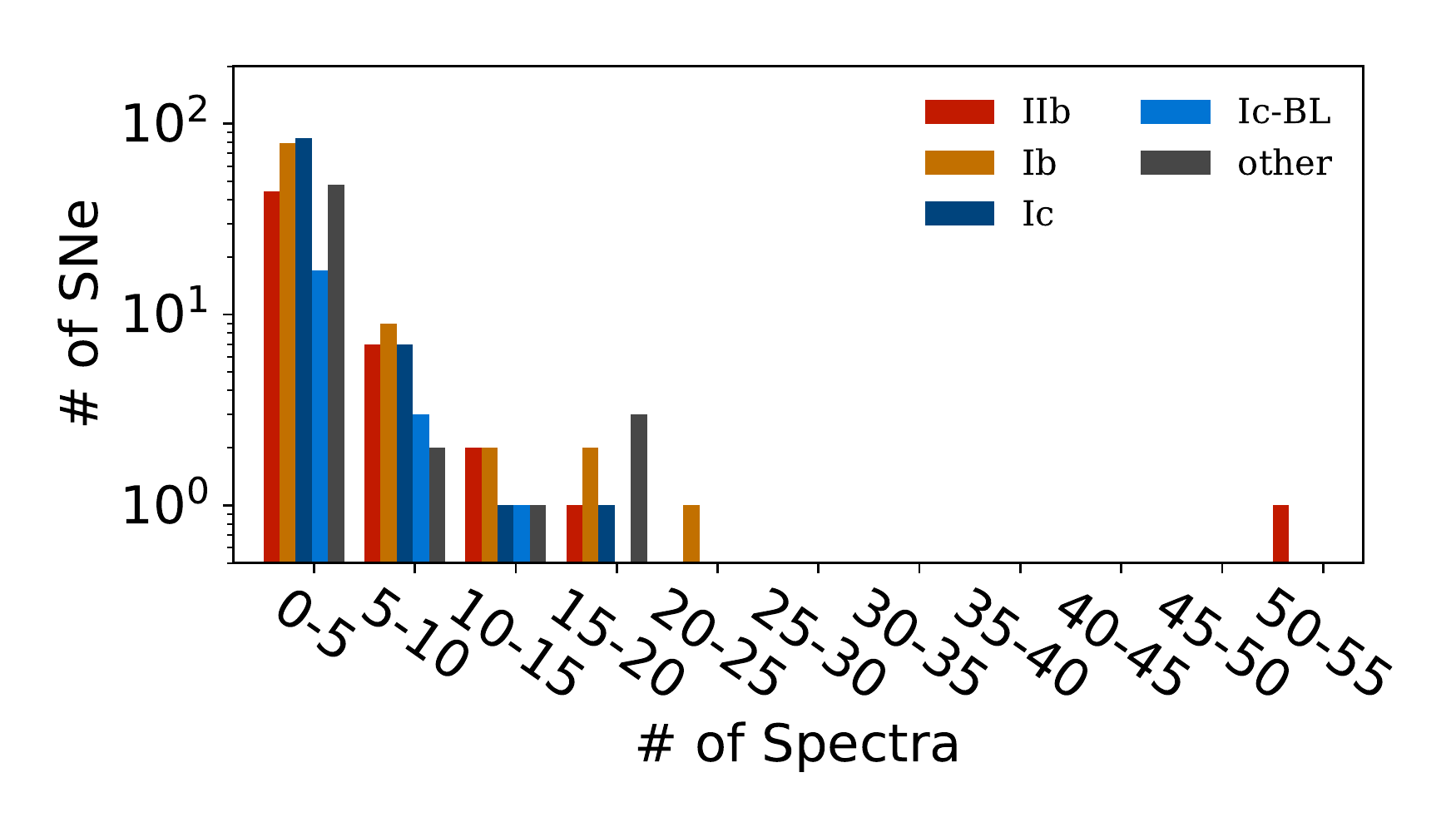}
    \caption{The distribution of spectral observation counts
    for SNe in our sample. Totals are binned over 5-spectra intervals,
    and we group spectra of peculiar subtypes and of stripped-envelope
    SNe without
    a clearcut subtype classification together into the ``other'' category.}
    \label{fig:spec_per_sn}
\end{figure}

\begin{figure}
    \includegraphics[width=0.45\textwidth]{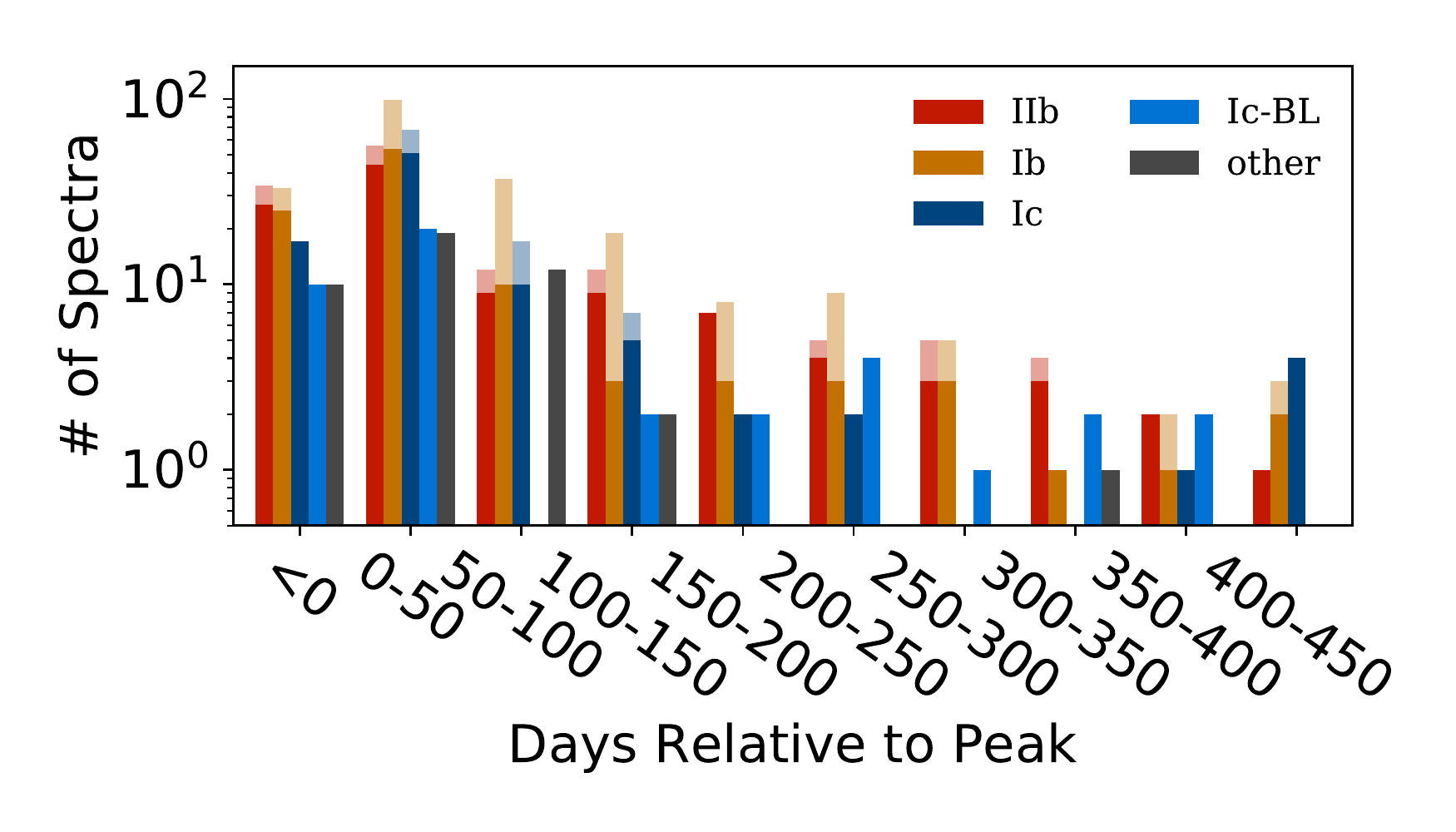}
    \caption{The phase distribution of spectra having a measured date of peak brightness,
grouped by spectral type and binned over 50-day intervals.
Spectra having photometric peak dates
are shown in solid, while the lighter shade also includes those
with {\tt SNID}-determined peak dates.  This plot is truncated at 450\,days;
a small number of observations exist at later phases.}
    \label{fig:phase_coverage}
\end{figure}

The majority of SNe in this sample have only a few spectral observations each,
though several well-observed examples of each subtype are also included ---
Figure~\ref{fig:spec_per_sn} shows the distribution of observation count
across SNe in our sample.
Our sample includes spectra taken within days of core collapse out to several years
(for those SNe that show ongoing, luminous interaction with circumstellar material (CSM)).
Figure~\ref{fig:phase_coverage} illustrates the phase coverage for spectra
of those SNe having either a photometrically- or {\tt SNID}-measured peak-brightness date.

\section{Results}
\label{sec:results}

Below we examine the evolution, relative to date of peak brightness, of several different observational
signatures in order to probe the kinematics and composition of the ejecta and to compare to previous
results.  In summary, we confirm that the
\ion{O}{1}\,$\lambda$7774 feature has both a higher pseudo-equivalent width (pEW) and a higher
characteristic velocity in SNe~Ic than in SNe~Ib or IIb, supporting the interpretation that SNe~Ic are stripped
further down toward the C-O core than SNe~Ib/IIb but they explode with similar energies.
However, we do not find supporting evidence for that scenario from an analysis of the nebular spectra:
one may expect that SNe~Ic, having lost more of their envelope but exploding at similar energies,
would exhibit wider full width at half-maximum intensity (FWHM) measures at very late times than do the SNe~Ib/IIb,
which we do not see.  We also highlight a few unique characteristics of this dataset, including the
large number of late-time spectra and the many spectra of rare SN subtypes released here.

\subsection{Oxygen Absorption}
\label{sec:oi}

\ion{O}{1}\,$\lambda$7774 absorption is readily apparent in the spectra of all three
major stripped-envelope classes, and it provides several important insights into the underlying
progenitor differences that give rise to these observational classes of SNe.
We measure two characteristics of this oxygen absorption feature in our spectra:
to compare line strength across SNe we calculate their pseudo-equivalent widths,
and to compare ejecta velocities we calculate the relativistic Doppler velocity indicated
by the wavelength of each absorption feature's minimum ($\lambda_{\rm{min}}$).
To obtain these measures, we follow the procedures outlined by \citet{2012MNRAS.425.1819S}
using a set of supervised algorithms, and we visually inspect each measure to identify and
correct errors.

\citet{2001AJ....121.1648M} find the \ion{O}{1}\,$\lambda$7774 absorption feature
to be stronger in SNe Ic than SNe Ib and suggest a trend: that the strength of
\ion{O}{1}\,$\lambda$7774 absorption is inversely correlated with the degree of stripping
experienced by the progenitor before core collapse. This finding was corroborated by \citet{2016ApJ...827...90L}
\citet{2018arXiv180700100F}, and is confirmed again when analyzing the dataset presented here.

\begin{figure}
    \includegraphics[width=0.45\textwidth]{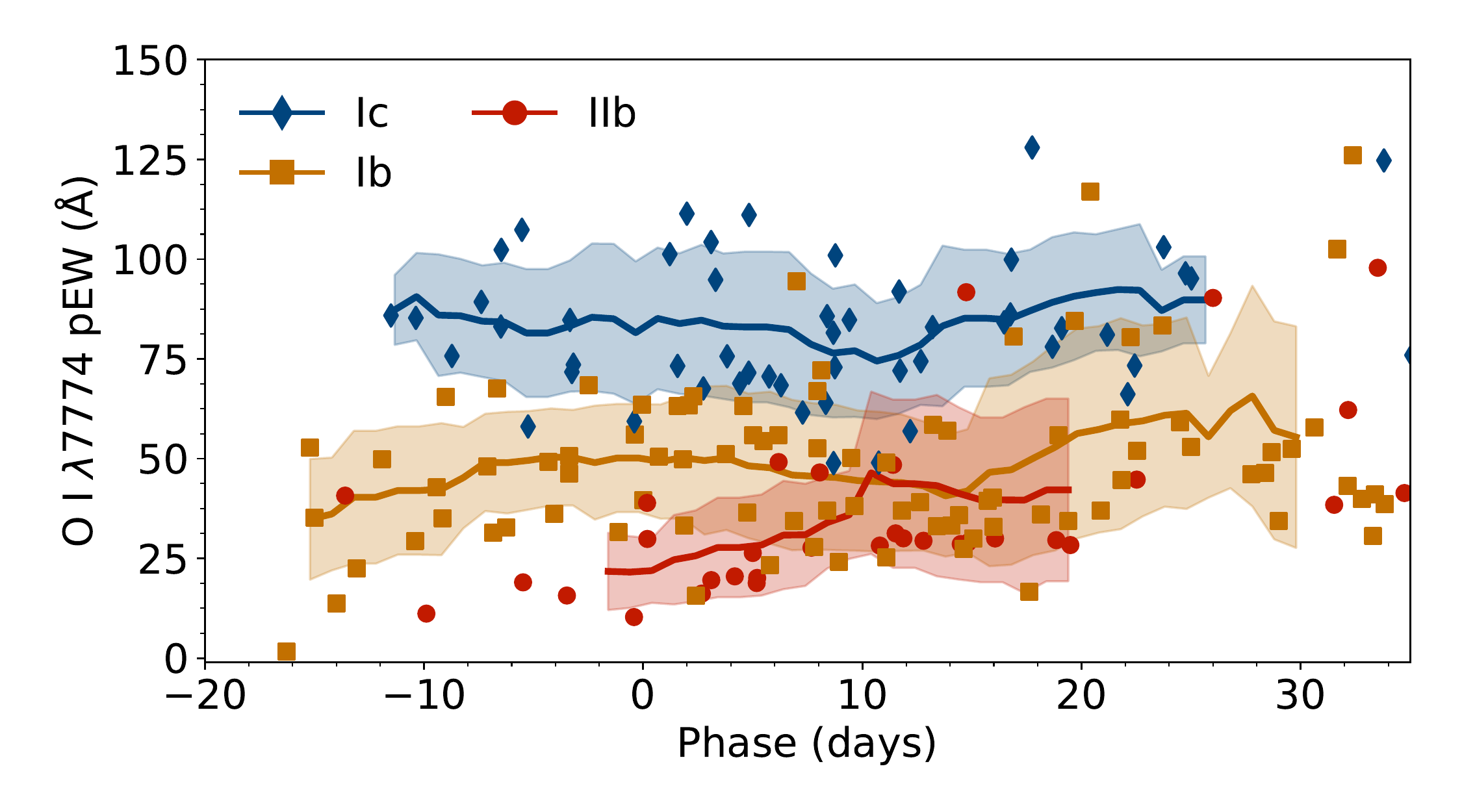}
    \caption{\ion{O}{1}\,$\lambda$7774 pEW as a function of phase for SNe~IIb, Ib, and Ic.
Measures from individual spectra are shown as points, with a rolling window mean and
associated standard deviations, calculated with a window size of 30\,days,
shown as lines and shaded regions (respectively).}
    \label{fig:OIpEW}
\end{figure}

Figure~\ref{fig:OIpEW} shows the pEW measures for the SNe~IIb, Ib, and Ic spectra
in this dataset having both a known phase and a clearly-identifiable \ion{O}{1}\,$\lambda$7774 feature,
along with a rolling window mean and the associated standard deviations. (Note that for this and all
following plots showing rolling window statistics, we ensure that no more than one measure
per SN is shown in each bin, so as to focus on the variance between events.)
At these phases, SNe~Ic exhibit significantly stronger oxygen absorption than
SNe~Ib and IIb, and there appears to be a small but systematic
difference between the line strengths in SNe~Ib and SNe~IIb, in good agreement with
the scenario put forth by \citet{2001AJ....121.1648M}.

\begin{figure}
    \includegraphics[width=0.45\textwidth]{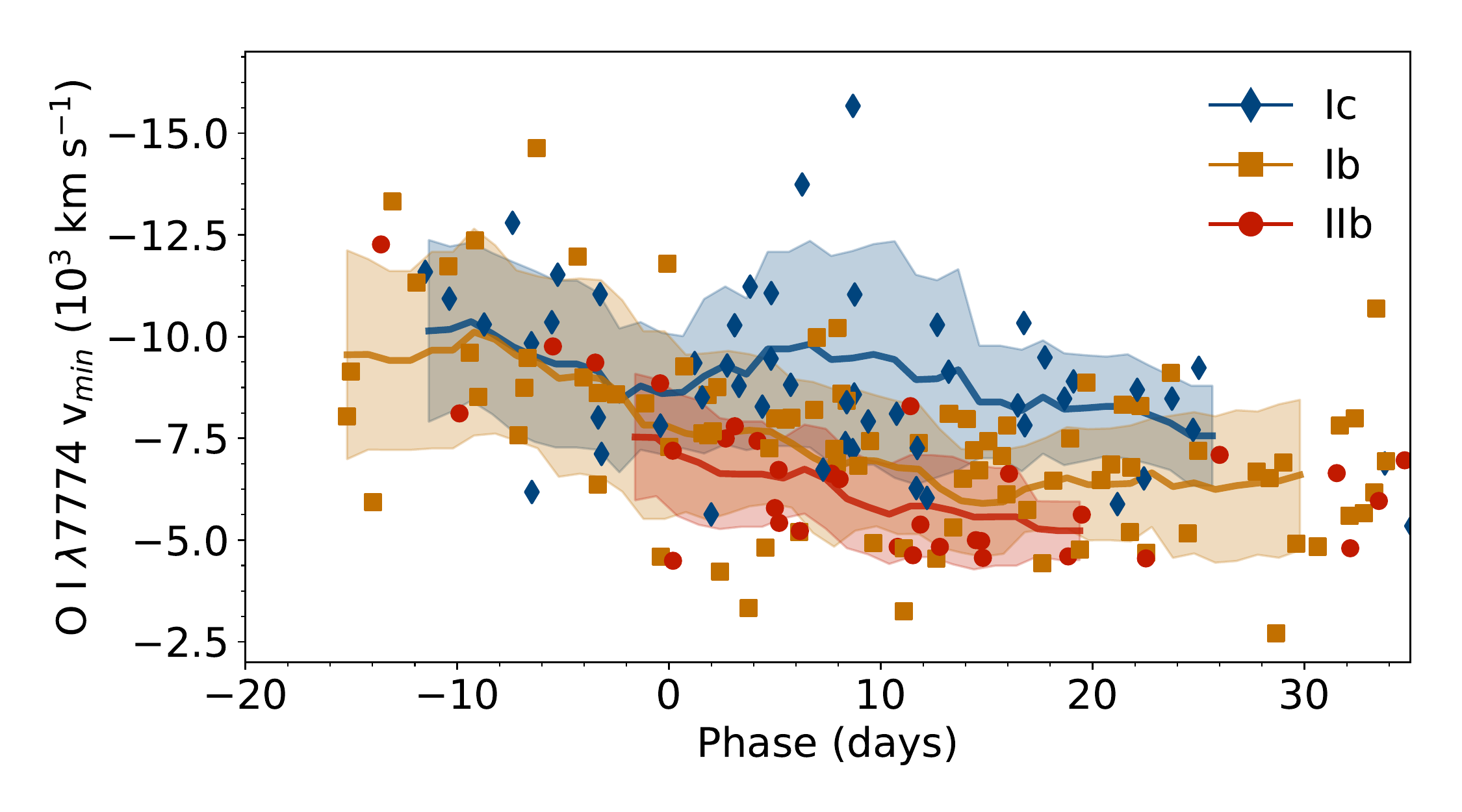}
    \caption{Velocity of \ion{O}{1}\,$\lambda$7774 absorption minimimum as a function of phase
for SNe IIb, Ib, and Ic.  Measures from individual spectra are shown as points,
with a rolling window mean and associated standard deviations, calculated with a window size of 30\,days,
shown as lines and shaded regions.}
    \label{fig:OIvmin}
\end{figure}

\citet{2001AJ....121.1648M}, \citet{2016ApJ...827...90L}, and \citet{2018arXiv180700100F} find
faster expansion velocities in SNe~Ic when compared to SNe~Ib, again based upon
the \ion{O}{1}\,$\lambda$7774 absorption feature.  Our data confirm those
results; see Figure~\ref{fig:OIvmin}.  While SNe~IIb, Ib, and Ic all show very similar
expansion velocities before peak brightness, in the $\sim 30$ days after peak they diverge somewhat
and exhibit a clear trend with $v_{\rm{Ic}} > v_{\rm{Ib}} \geq v_{\rm{IIb}}$ before settling
back into a single distribution at 30+\,days.

\subsection{Nebular Emission Lines}
\label{sec:nebular_emission}

\citet{2001AJ....121.1648M} examine the FWHM
values of forbidden emission features in late-time stripped-envelope SN spectra.
They find SNe Ic to have broader emission than SNe~Ib,
and they propose this to be a result of similar explosion energies
but varying helium-hydrogen envelope masses across the
subclasses at the time of core collapse.
This data release includes a much larger set of late-time spectra ---
we include some 200+ spectra having a well-determined phase
(see \S\S\ref{sec:peaks}, \ref{sec:snidpeaks})
and observed 60+ days after peak.

\begin{figure}
    \includegraphics[width=0.45\textwidth]{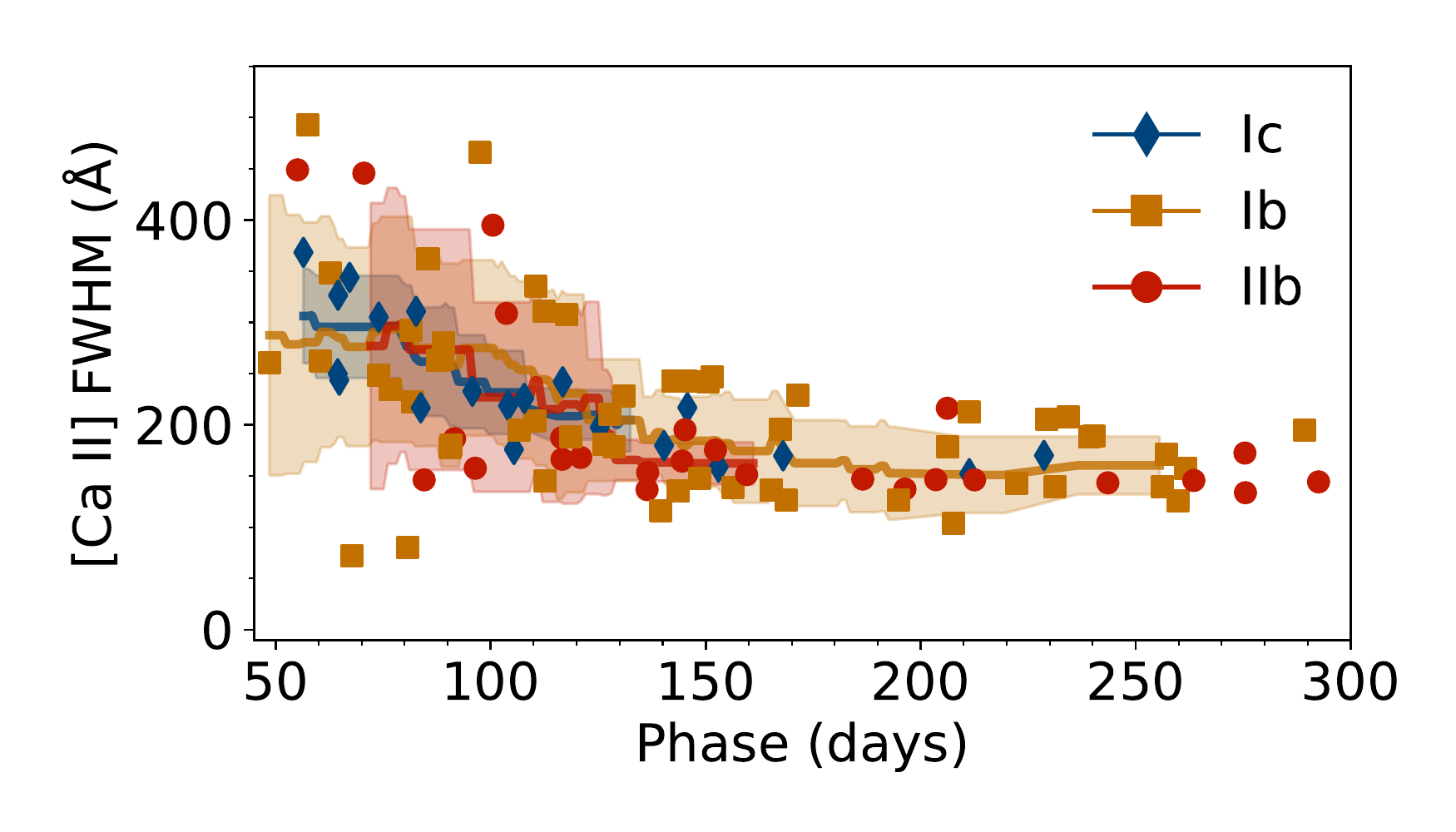}
    \caption{The FWHM of the blended [\ion{Ca}{2}]\,$\lambda\lambda$7319, 7324 doublet as
    a function of phase for SNe~IIb, Ib, and Ic.  Lines and shaded regions show
    a rolling window mean and standard deviation with window size of 50\,days.
    We find that the three distributions overlie each other
    with no significant differences.}
    \label{fig:nebular_widths}
\end{figure}

To compare, we measure the FWHM of the
[\ion{Ca}{2}]\,$\lambda\lambda$7319, 7324 blended doublet, which
is strong and relatively well isolated from other emission features for
all subtypes of stripped-envelope SNe.
(Note that this feature arises from a doublet, but only a small fraction
of the observed line widths can be attributed to the 5\,\AA\ ($\sim200$\,\kms)
separation between the doublet.)
We find no evidence for a systematic difference in late-time emission-line widths
between SNe IIb, Ib, and Ic; see Figure~\ref{fig:nebular_widths}.
Between 50 and 100\,days post-peak, the measured widths for all
subtypes show a wide scatter with an overall trend downward, as
the SNe are transitioning into the nebular phase and the forbidden calcium
emission begins to dominate the measurements, but after day 100 all three
subtypes exhibit little evolution and hover around
$\rm{FWHM} \approx 200$\,\AA\ ($8200$\,\kms).
This is in good agreement with the $8700 \pm 2700$\,\kms\ that \citet{2001AJ....121.1648M}
found for the 9 SNe Ic spectra in their sample, but is in disagreement with the
$4900 \pm 800$\,\kms\ they calculated from their small sample of four SNe Ib.

\subsection{Helium-Strength Evolution}
\label{sec:helium_evolution}

\begin{figure}
    \includegraphics[width=0.45\textwidth]{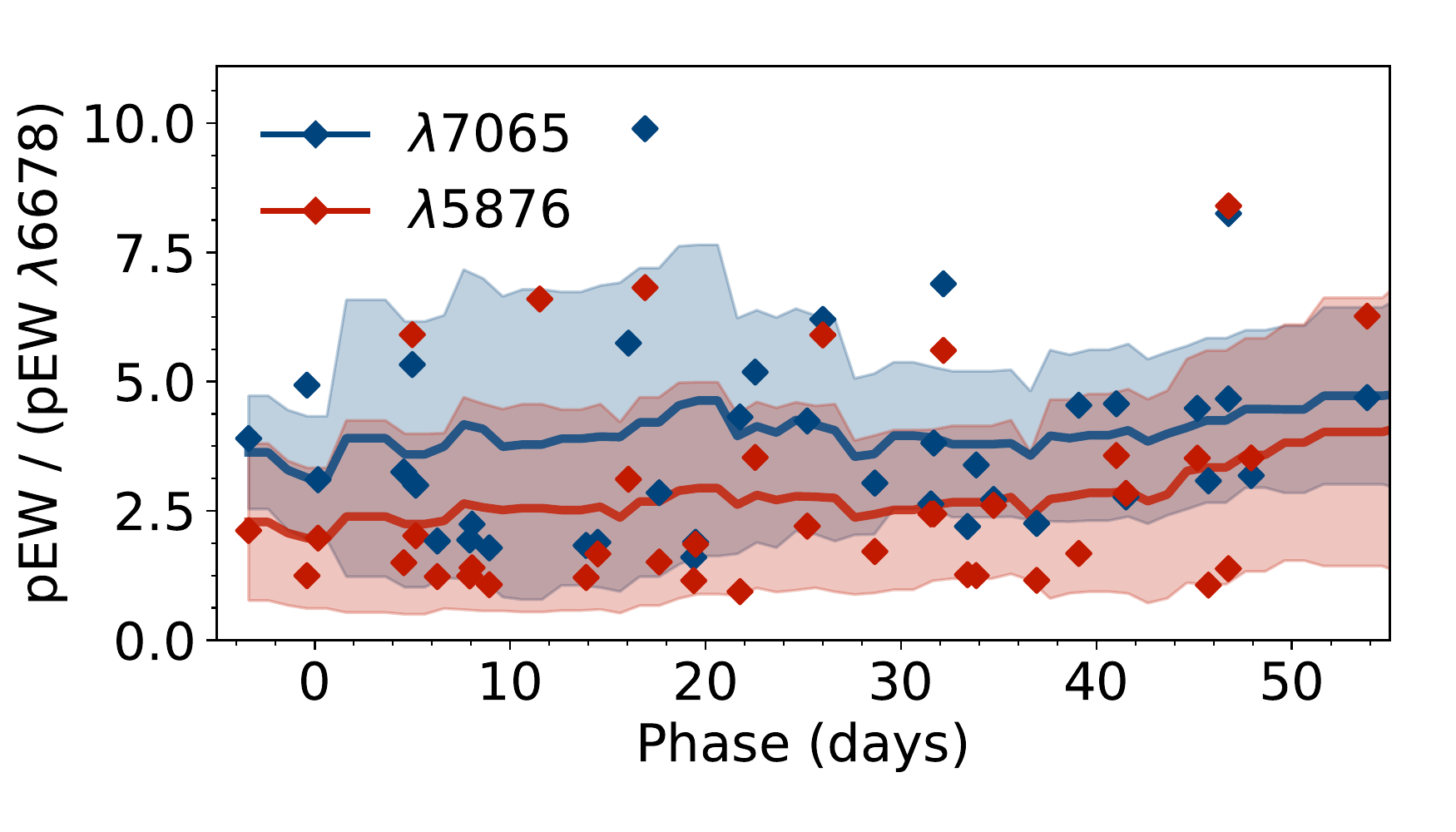}
    \caption{The pEW of \ion{He}{1}\,$\lambda$5876 and \ion{He}{1}\,$\lambda$7065
    relative to that of \ion{He}{1}\,$\lambda$6678, in the SN Ib and SN IIb spectra
    having a known phase and robust detections of all three lines.
    The rolling window mean and standard deviation, calculated with a window size of 20\,days,
    is shown with lines and shaded regions (respectively).
    The relative strengths of these three lines are consistently
    ordered $\rm{pEW}_{\lambda5876} > \rm{pEW}_{\lambda7065} > \rm{pEW}_{\lambda6678}$,
    but we find no significant temporal evolution in these ratios.}
    \label{fig:he_i_pew}
\end{figure}

\citet{2001AJ....121.1648M} find evidence for a temporal evolution in the relative
strengths of helium lines in SNe~Ib, with both \ion{He}{1}\,$\lambda$5876
and \ion{He}{1}\,$\lambda$7065 growing in strength compared to \ion{He}{1}\,$\lambda$6678
after day $\sim30$. \citet{2016ApJ...827...90L}, however, do not find such a trend
in their pEW measures, and the data presented here do not indicate a significant trend.
Figure~\ref{fig:he_i_pew} shows the relative pEWs of these three \ion{He}{1} lines,
normalised to the \ion{He}{1}\,$\lambda$6678 measure, for those SN Ib and IIb
spectra having a known phase and clear simultaneous detections of all three lines.
We also examine these line ratios amongst the SN Ib and SN IIb populations separately, and
again find no evidence for strong temporal evolution.

\subsection{Mean Spectra}
\label{sec:meanspec}

\begin{figure*}
    \centering
    \includegraphics[width=0.49\textwidth]{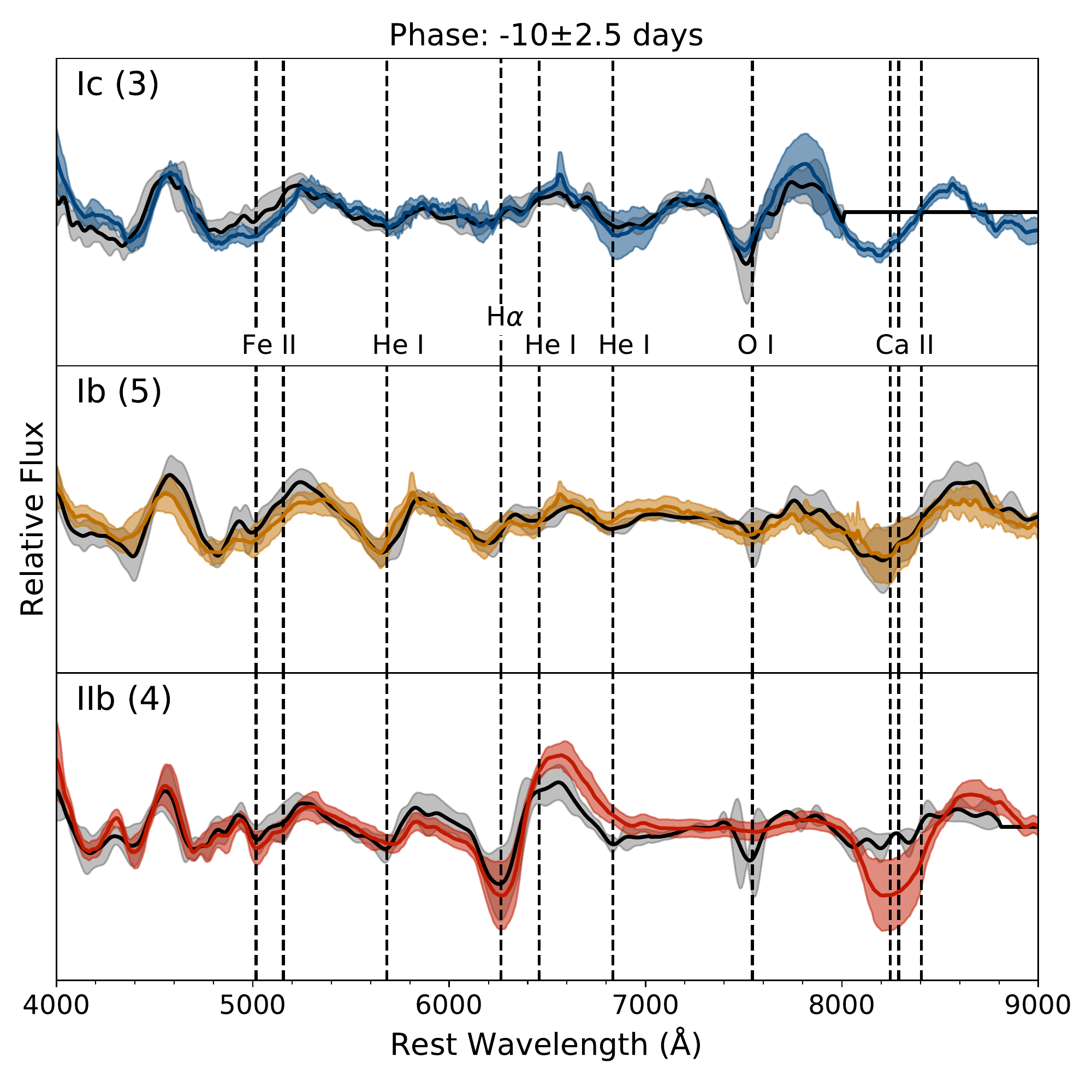}
    \includegraphics[width=0.49\textwidth]{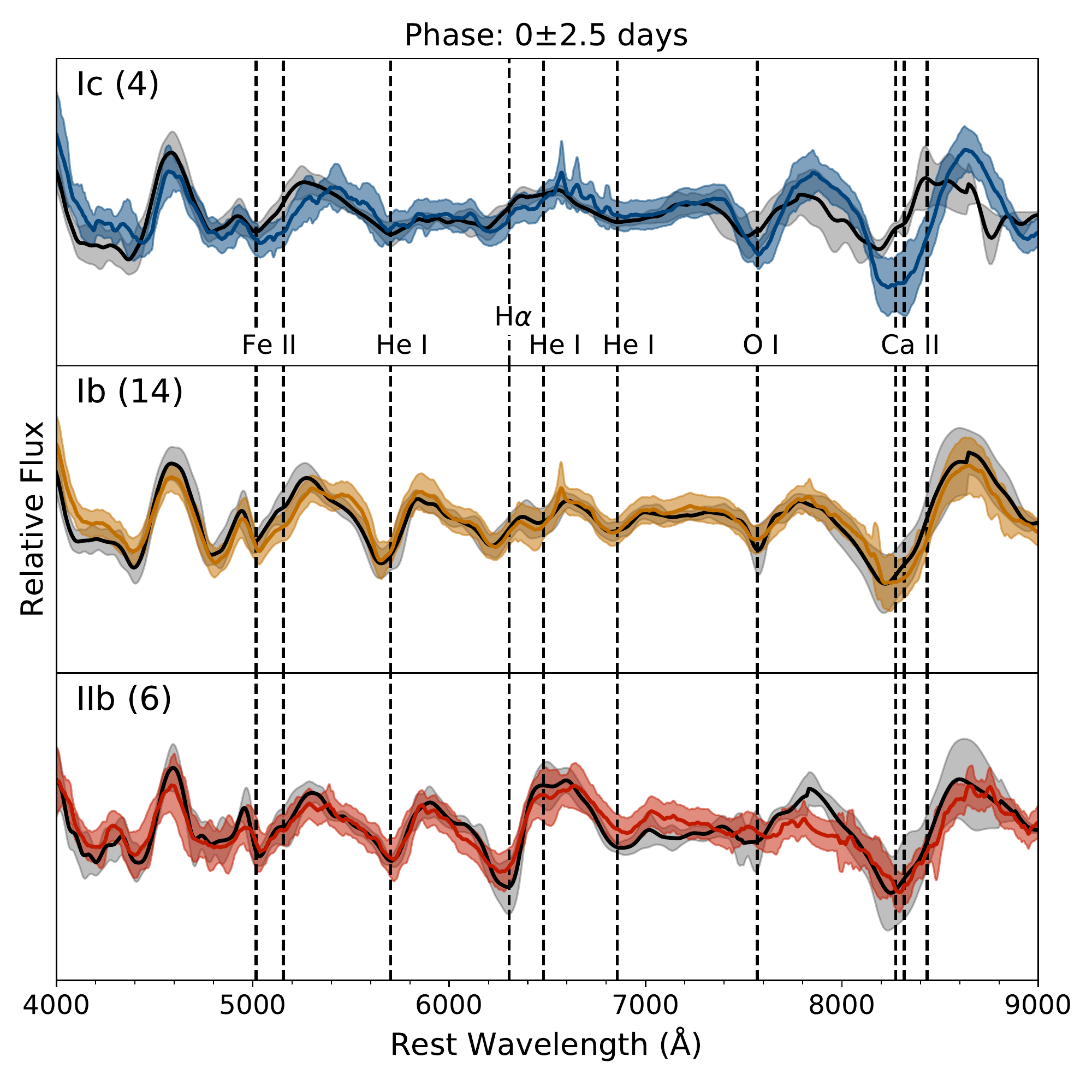}
    \includegraphics[width=0.49\textwidth]{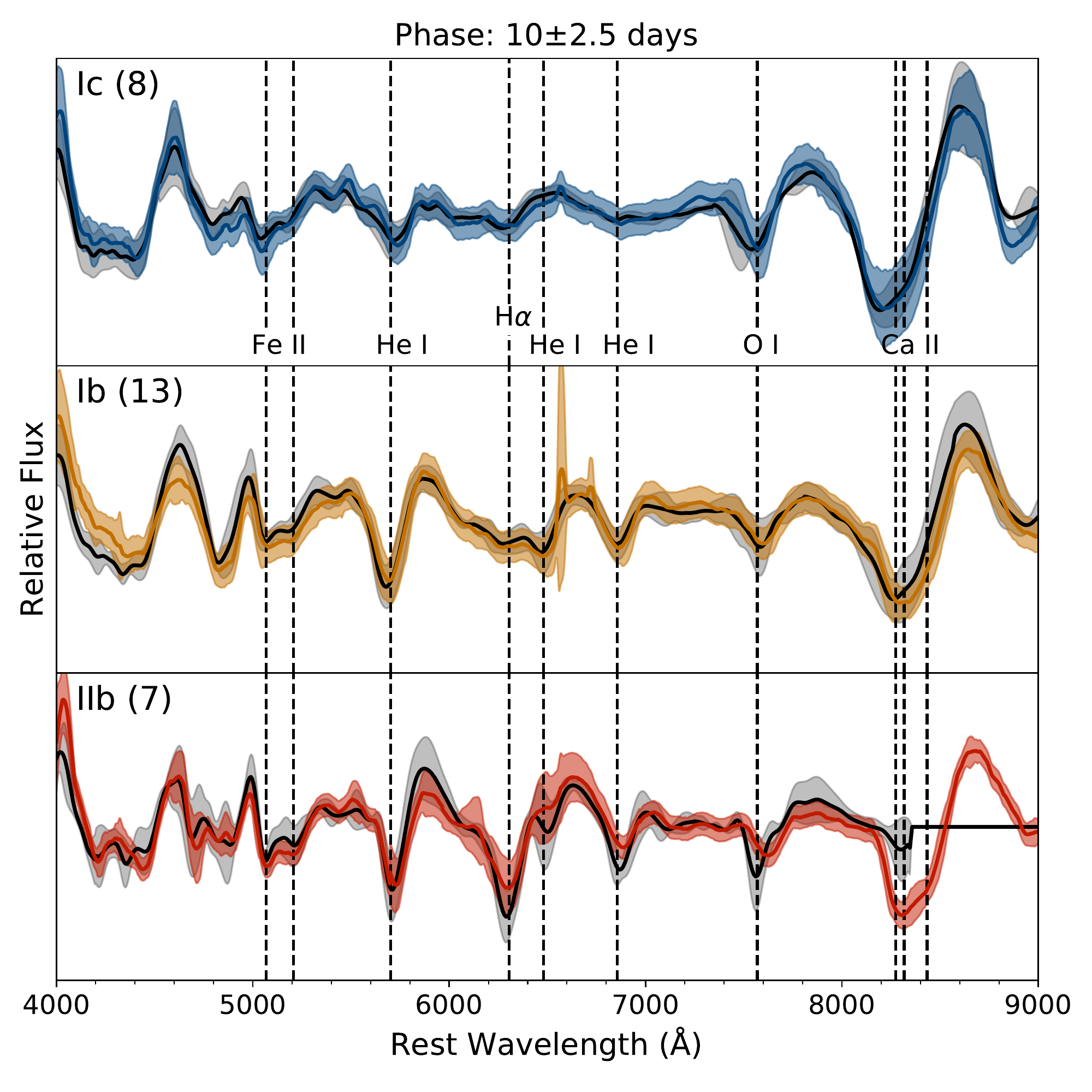}
    \includegraphics[width=0.49\textwidth]{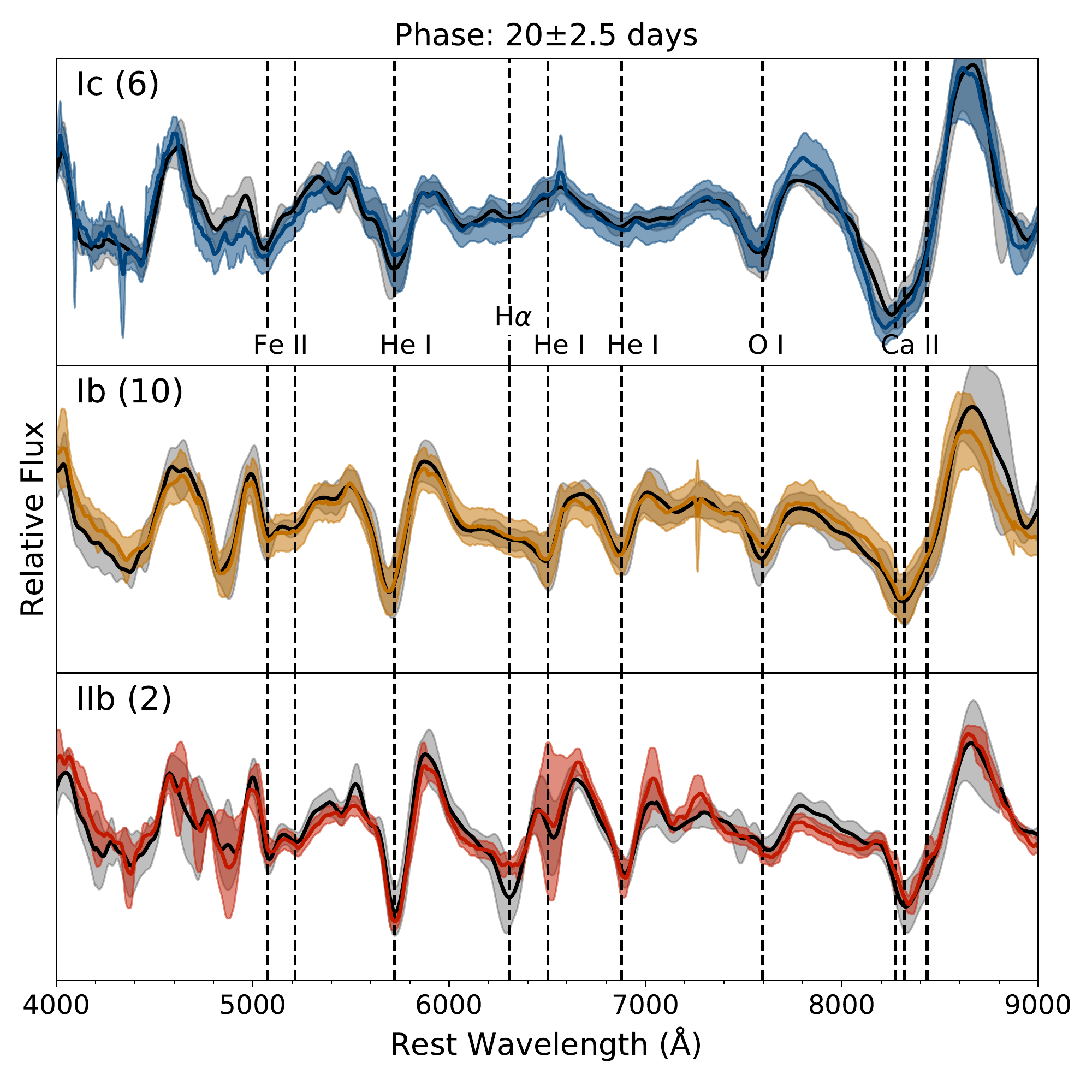}
    \caption{Mean spectra calculated in four different phase bins for normal members
of the three dominant stripped-envelope SN subtypes.
Within each bin, we include only a single spectrum per SN, and the
number of SNe represented by each panel is shown in parentheses.
Spectra calculated from our sample are shown in colour and those
presented by \citet{2016ApJ...827...90L} are in grey, for comparision.
Major line identifications are provided (though note that not all of these
lines are visible in all SN types), and the standard deviation within the bin
is illustrated by the shaded regions.}
    \label{fig:meanspec}
\end{figure*}

To illustrate the spectral line variations between major subclasses over time,
we construct continuum-removed mean spectra at a few phases (relative to peak)
for each major spectral subtype,
in much the same manner as \citet{2016ApJ...827...90L} and \citet{2017ApJ...845...85L}.
In detail, this process involves several steps. First we take all SNe within our sample that are
classified as normal SNe IIb, Ib, or Ic and which have a date of peak brightness measurement
as described in \S\ref{sec:peaks} or \S\ref{sec:snidpeaks}.  We take
those spectra observed within four different phase bins
(each bin has a width of $\pm 2.5$\,days and they are centered at $-10$, 0, $+10$, and $+20$ days relative
to peak brightness)
and choose a single spectrum per SN per bin, preferring spectra that span the entire
wavelength range and those observed at a phase nearest the bin center.

After applying redshift and Milky Way dust reddening corrections, we fit for a smooth
pseudo-continuum.  We then approximate each spectrum's effective continuum with a cubic spline
and we normalise the spectra by dividing each by their continuum,
to allow line-strength comparisons across different SNe (which are likely affected
by different amounts of host-galaxy dust reddening).
We smooth the continuum-divided spectra via convolution with a window of width 21\,\AA\
and interpolate onto a single shared wavelength array.  Within each binned set
of SNe, we then find the mean value and standard deviation of the normalised
spectra as a function of wavelength.  Our results are shown in
Figure~\ref{fig:meanspec}.

The key spectral differences between classes are readily apparent in Figure~\ref{fig:meanspec}.
Strong H$\alpha$ is visible in the SN~IIb spectra at all phases, but
not in the SN Ib and SN Ic spectra.
Strong \ion{He}{1} lines are found in the SN Ib and SN IIb spectra, but not in the SN Ic spectra.
\ion{O}{1} is found in all, but is notably stronger in the SN Ic spectra than in the SN Ib and SN~IIb spectra
(see \S\ref{sec:oi}). \ion{Fe}{2} features can be found toward the blue
edge for all SN types, and the near-infrared \ion{Ca}{2} triplet dominates their
red extremes.

Previous authors \citep[e.g.,][]{2016ApJ...820...75P} have argued for the presence
of weak high-velocity H$\alpha$ in some SNe Ib and Ic,
and \citet{2016ApJ...827...90L} find a high-velocity feature in their mean spectra
at $-10$ and 0\,days that is reasonably interpreted to be this high-velocity hydrogen.
The same feature is apparent in our SN Ib mean spectra at $-10$ and 0 days,
just blueward of the H$\alpha$ feature marked on the SN IIb spectra.

\subsection{Nebular-Phase Spectra}
\label{sec:nebular}

At late times ($\gtrsim 60$\,days), the ejecta of stripped-envelope SNe
become largely transparent and transition into the nebular phase.
This data release includes a large set of spectra observed after this transition.
These spectra do not show a continuum flux level
against which to normalise out the effects of host-galaxy dust
obscuration, as we have done with the photospheric spectra
in \S\ref{sec:meanspec}, so it is difficult to perform
quantitative analyses of the same sort across the entire set
of observations.

\begin{figure}
    \includegraphics[width=0.5\textwidth]{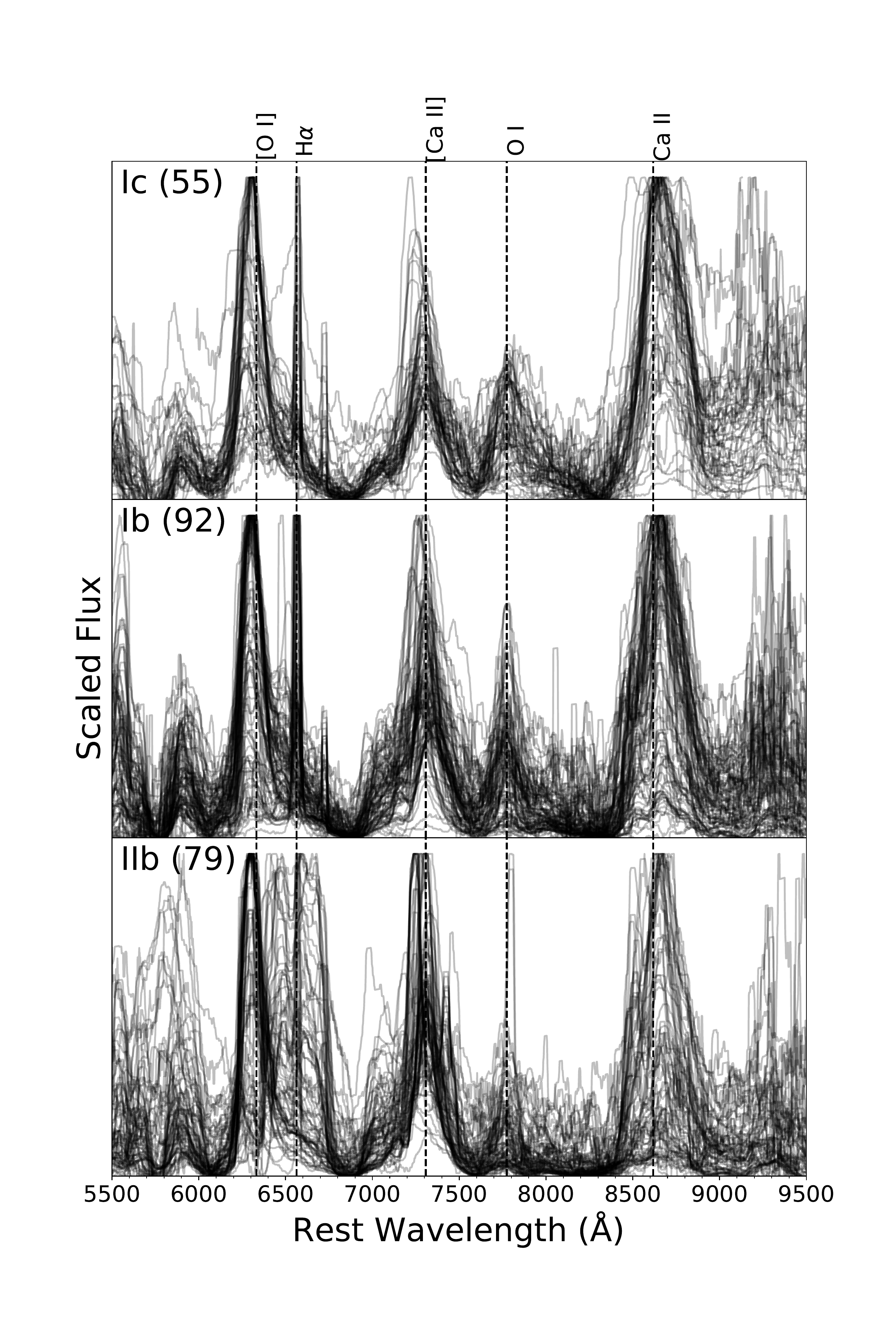}
    \caption{Nebular spectra of SNe~Ic, Ib, and IIb are shown
    in the top, middle, and bottom panels (respectively).
    The number of spectra is listed in parentheses, and each
    spectrum has had a linear continuum subtracted and has been
    renormalised to range between 0 and 1. This format has been
    chosen for display purposes; the plotted minima of these spectra
    may differ from zero flux.}
    \label{fig:nebular}
\end{figure}

Instead, in Figure~\ref{fig:nebular} we simply plot these
nebular spectra overlain atop each other as an illustration of
the qualitative trends in the data.
It is apparent that nebular-phase spectra of the three main
stripped-envelope SN subclasses are quite similar, and are dominated by
the same oxygen and calcium emission features having similar strengths
and line widths.  While the only H$\alpha$ emission found in
the SN~Ic and SN~Ib spectra is the occasional unresolved galactic emission line,
a few of the SN~IIb spectra in our sample show broad nebular
H$\alpha$ emission associated with ongoing interaction with
extended H-rich CSM \citep[e.g.,][]{2000AJ....120.1487M}.

Figure~\ref{fig:nebular} may suggest a few other trends as well,
though we again caution the reader that the spectra in these plots have
not been corrected for host-galaxy dust obscuration, and they have
been independently renormalised; we leave a detailed exploration of
the nebular-phase spectra to future work.

\subsection{Subpopulations of Interest}
\label{sec:rare}

Stripped-envelope SNe represent a sizeable fraction of all SNe in a volume-limited sample
\citep[$\sim 31$\,\% of core-collapse SNe are stripped-envelope; e.g.,][]{2011MNRAS.412.1441L,2011MNRAS.412.1522S,2017PASP..129e4201S},
with an overall rate in good agreement with the observed population fraction of their most common progenitors
\citep[$\sim 33$\,\% of massive stars lose their outer envelope via binary interaction before core collapse; e.g.,][]{2012Sci...337..444S}.
In addition to the ``normal'' SNe~IIb, Ib, and Ic, a diverse but relatively small set of outliers
has also been discovered in the stripped-envelope SN population, exhibiting
anomalous observed spectral properties, light curves, luminosities, and explosion-site locations.

Some of these outliers appear to be (in some way or another) unique amongst the as-yet-observed
set of SNe, but many others are well clustered in their characteristics and form interesting subpopulations,
including the superluminous SNe~Ic (SLSNe~Ic), the SN~Ib-like SNe with strong \ion{Ca}{2} lines relative to \ion{O}{1} lines
(the ``Ca-rich'' SNe), and the SNe~Ib and IIb showing evidence of interaction with dense CSM
(the SNe~Ibn and IIbn).
The progenitor channels for these relatively rare subpopulations are only poorly understood, if at all,
and many puzzles about their origins remain.  Here we highlight the rare subpopulation observations published
as part of this sample.  These subtypes are generally not well-represented in the {\tt SNID} template set,
and so the spectral classification scheme described in \ref{sec:classification} is not adequate --- these
examples were discovered and classified by hand through various methods, usually after the automatic classification
workflow used for the normal events had failed, and it is difficult to assess how complete our classifications are.

\begin{figure}
    \includegraphics[width=0.45\textwidth]{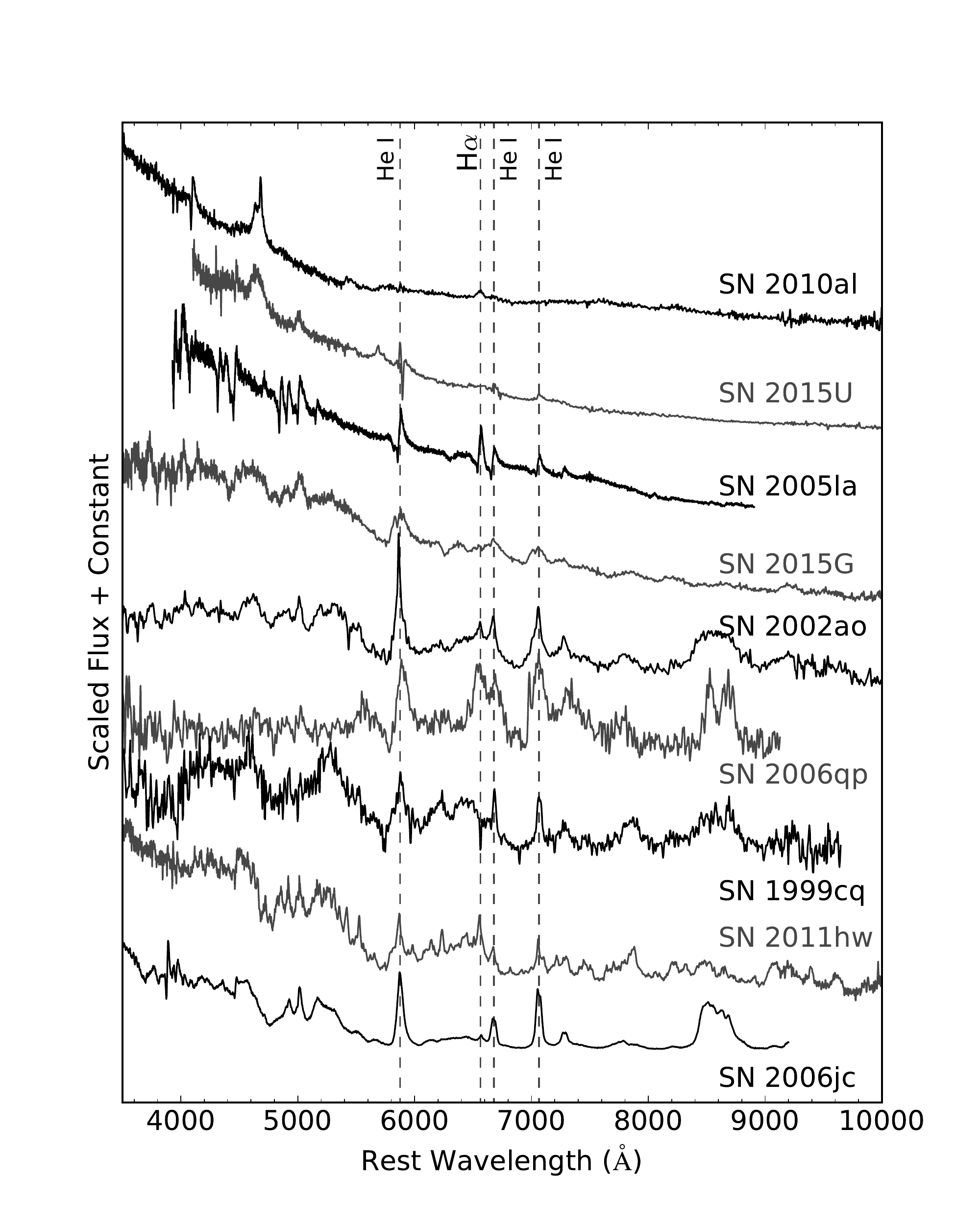}
    \caption{Spectra of the SNe Ibn/IIbn included in this dataset.
    Subsets of these spectra were previously published by
    \citet{2016MNRAS.461.3057S,2017MNRAS.471.4381S}, \citet{2007ApJ...657L.105F},
    \citet{2000AJ....119.2303M}, and \citet{,2012MNRAS.426.1905S}.
}
    \label{fig:ibns}
\end{figure}

A small subset of events exhibit the properties associated with the presence of very dense
CSM, most notably narrow spectral emission lines arising from the ionised but not yet shock-accelerated
material.  These events have been dubbed SNe Ibn (``n'' for ``narrow lines''), following the nomenclature used for the
more common SNe IIn \citep[e.g.,][]{1990MNRAS.244..269S,1991ESOC...37..343F,2000AJ....119.2303M,2007Natur.447..829P}.
There have also been a few events that show only very weak hydrogen features, and they have been
labeled SNe IIbn \citep[e.g.,][]{2015MNRAS.449.1921P}.
Observations of several of the well-studied members of this subclass are included in this sample
\citep[much of which has already been published; e.g., SNe~2006jc and 2015G;][]{2007ApJ...657L.105F,2016MNRAS.461.3057S,2017MNRAS.471.4381S},
but we also include in this release several observations published here for the first time; see Figure~\ref{fig:ibns}.

\begin{figure}
    \includegraphics[width=0.45\textwidth]{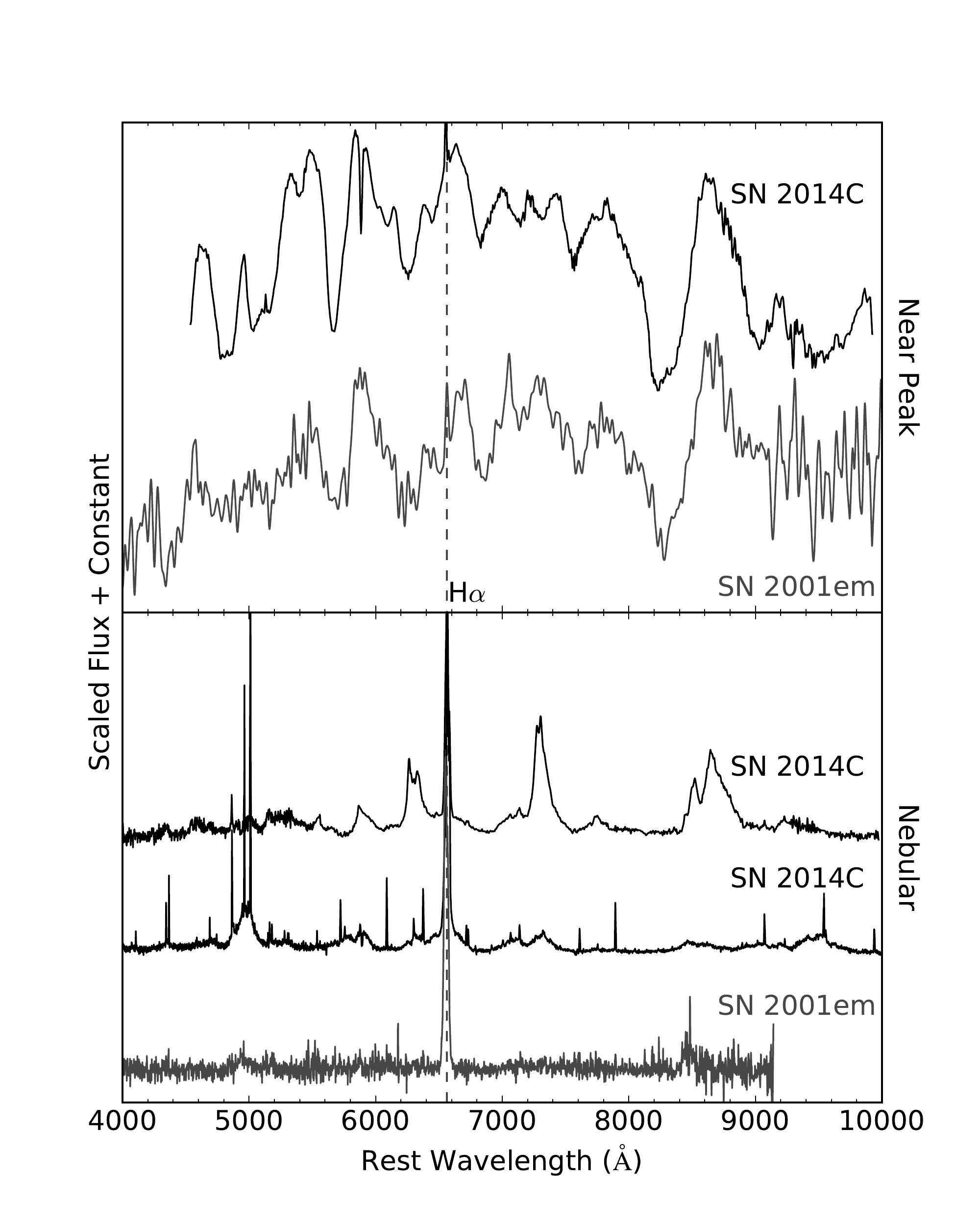}
    \caption{Spectra of the late-interacting SNe~Ib included in this release.
    A subset of these spectra was previously published by \citet{2018MNRAS.tmp.1325M}.}
    \label{fig:late-interactors}
\end{figure}

In addition to the SNe Ibn/IIbn, a very small set of events have appeared to be more-or-less
normal SNe Ib in their spectral properties at peak brightness, but then showed strong interaction-driven hydrogen
emission features at late phases
\citep[SNe 2001em, 2004dk, and 2014C; e.g.,][]{2006ApJ...641.1051C,2015ApJ...815..120M,2017ApJ...835..140M,2018MNRAS.tmp.1325M}.
These events likely exploded inside the cavity of a dense shell of CSM lost from their
progenitors in the few thousands of years before core collapse, and several examples of
SNe having hydrogen-rich ejecta but similar shells of CSM have been found \citep[e.g., SN~1996cr;][]{2008ApJ...688.1210B}.
Though we present no new events of this nature, included in this data release are several
newly-published observations of both SN~2001em and SN~2014C; see Figure~\ref{fig:late-interactors}.

\begin{figure}
    \includegraphics[width=0.45\textwidth]{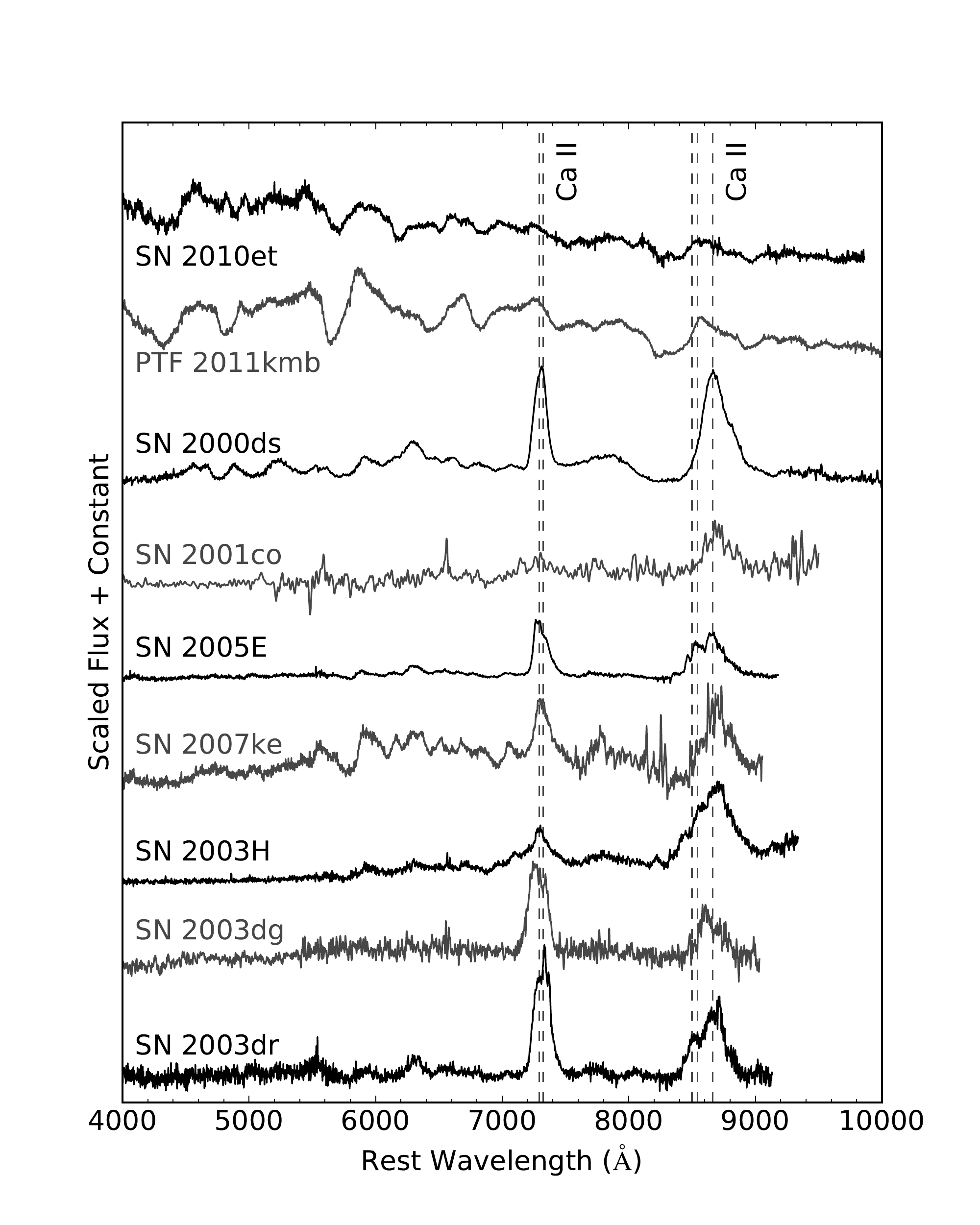}
    \caption{The Ca-rich SNe included in this dataset.  Spectra taken near
    peak brightness are plotted for SN~2010et and PTF~2011kmb, while the others
    were taken well past peak at nebular phases.
    Subsets of these spectra were previously published by
    \citet{2012ApJ...755..161K}, \citet{2009AJ....138..376F}, and \citet{2015MNRAS.452.2463F}.}
    \label{fig:ca-rich}
\end{figure}

There is a population of events that spectroscopically appear to be of Type Ib at maximum brightness,
yet they have relatively low luminosity, fade rapidly, exhibit peculiarly strong nebular
\ion{Ca}{2} lines at late phases, and explode far from the strongly star-forming regions
with which core-collapse events are generally associated -- the ``Ca-rich'' SNe
\citep{2003IAUC.8159....2F,2010Natur.465..322P}.
The exact progenitor channel for this subclass remains uncertain \citep{2012ApJ...755..161K,2015MNRAS.452.2463F};
Figure~\ref{fig:ca-rich} shows the Ca-rich events with data included in this release.

\begin{figure}
    \includegraphics[width=0.45\textwidth]{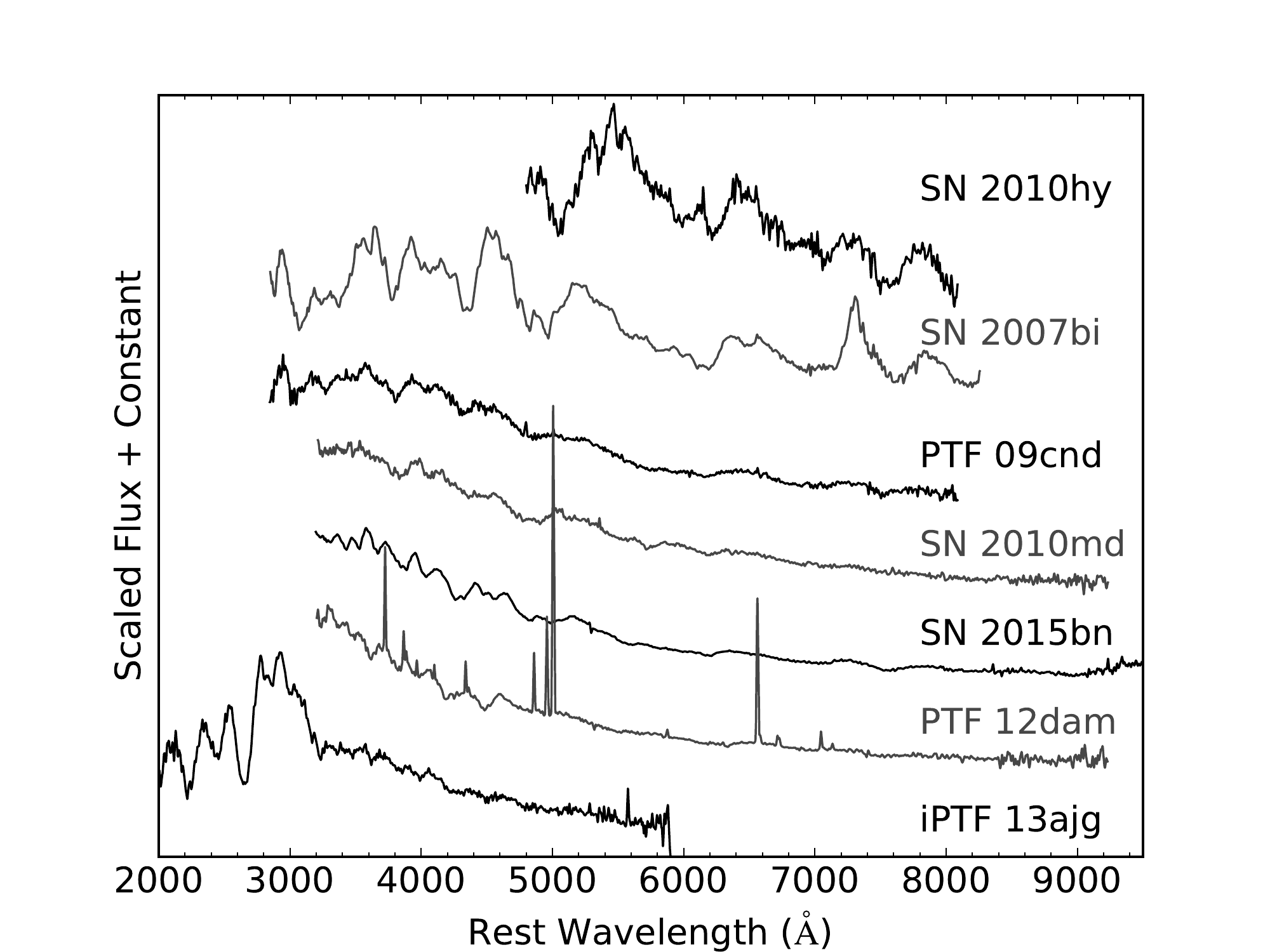}
    \caption{A few of the SLSN~I observations included in this release.
    Subsets of these spectra were previously published by
    \citet{2009Natur.462..624G} and \citet{2014ApJ...797...24V}.}
    \label{fig:slsn}
\end{figure}

Finally, we also make note of the few observations of SLSNe-I included in this release.
Superluminous SNe are generally an order of magnitude more luminous than normal core-collapse
SNe, and they come in both hydrogen-rich (SLSNe-II) and hydrogen-poor varieties
\citep[SLSNe-I; e.g.,][]{2012Sci...337..927G,2017ApJ...845...85L}.
Energy contributions from a newly-formed magnetar provide a plausible progenitor channel
for these brilliant events, but many questions still remain
\citep[e.g.,][]{2012MNRAS.426L..76D,2013ApJ...770..128I,2017ApJ...850...55N};
Figure~\ref{fig:slsn} shows a few of the SLSN~I observations included in this release.

\section{Conclusions}

We present the complete sample of stripped-envelope SN spectra observed by the
U.C.\ Berkeley SN group over the last 3 decades: 888 spectra of 302 SNe, including
384 spectra (of 92 SNe) with photometrically-determined phases.
Comparisons with previous work show that this dataset proves
a powerful addition to the observational study of massive stellar deaths.

We confirm previously-noted trends in the \ion{O}{1}\,$\lambda$7774 line properties
of these events: both the strength and characteristic velocity of oxygen absorption
are higher in SNe~Ic than in SNe~Ib and IIb at phases near peak brightness.
This is best understood as evidence that the progenitors of Type Ic SNe have had their
outer hydrogen and helium envelopes more heavily stripped than have those of Types
Ib or IIb, and that the spectra of SNe~Ic are probing more deeply into their
more strongly accelerated C-O cores.

However, contrary to previous claims and somewhat at odds with the above result,
we find that the line widths of late-time emission features are not a function of spectral type.
We also find that the relative strengths of \ion{He}{1} absorption lines in SNe~Ib and IIb
are not strongly correlated with phase, emphasizing that capturing the date of peak brightness
in the light curve is of great value.

Finally, we produce continuum-removed mean spectra of SNe~Ic, Ib, and IIb
\citep[which are consistent with those of][]{2016ApJ...827...90L},
to help define the spectral
characteristics of the subclasses,
and we present new observations of several categories of rare and poorly-understood events.
Our spectra and all associated data tables are made available online.

\section*{Acknowledgements}

We would like to thank
Brian J. Barris,
Peter Blanchard,
Joshua S. Bloom,
Bethany E. Cobb,
Alison Coil,
Louis-Benoit Desroches,
Andrea Gilbert,
Christopher V. Griffith,
Luis C. Ho,
Saurabh W. Jha,
Michael T. Kandrashoff,
Minkyu Kim,
Nicholas Lee,
Adam A. Miller,
Matthew R. Moore,
Aleksandir Morton,
Robin E. Mostardi,
Peter E. Nugent,
Marina S. Papenkova,
Sung Park,
Daniel A. Perley,
David Pooley,
Dovi Poznanski,
Adam G. Riess,
Brad Tucker,
Vivian U,
XiangGao Wang,
and
Xiaofeng Wang
for their assistance with some of the observations over the last three
decades.
We are grateful to the staffs at Lick and Keck Observatories for their
hard work in making the observations possible. Some of the data
presented herein were obtained at the W. M. Keck Observatory, which is
operated as a scientific partnership among the California Institute of
Technology, the University of California, and the National Aeronautics
and Space Administration (NASA); the observatory was made possible by
the generous financial support of the W. M. Keck Foundation. The
authors wish to recognise and acknowledge the very significant
cultural role and reverence that the summit of Maunakea has always had
within the indigenous Hawaiian community; we are most fortunate to
have the opportunity to conduct observations from this mountain.

Research at Lick Observatory is partially supported by a generous gift
from Google.
We also greatly appreciate contributions from
numerous individuals, including
Charles Baxter and Jinee Tao,
Firmin Berta,
Marc and Cristina Bensadoun,
Frank and Roberta Bliss,
Eliza Brown and Hal Candee,
Kathy Burck and Gilbert Montoya,
Alan and Jane Chew,
David and Linda Cornfield,
Michael Danylchuk,
Jim and Hildy DeFrisco,
William and Phyllis Draper,
Luke Ellis and Laura Sawczuk,
Jim Erbs and Shan Atkins,
Alan Eustace and Kathy Kwan,
David Friedberg,
Harvey Glasser,
Charles and Gretchen Gooding,
Alan Gould and Diane Tokugawa,
Thomas and Dana Grogan,
Alan and Gladys Hoefer,
Charles and Patricia Hunt,
Stephen and Catherine Imbler,
Adam and Rita Kablanian,
Roger and Jody Lawler,
Kenneth and Gloria Levy,
Peter Maier,
DuBose and Nancy Montgomery,
Rand Morimoto and Ana Henderson,
Sunil Nagaraj and Mary Katherine Stimmler,
Peter and Kristan Norvig,
James and Marie O'Brient,
Emilie and Doug Ogden,
Paul and Sandra Otellini,
Jeanne and Sanford Robertson,
Stanley and Miriam Schiffman,
Thomas and Alison Schneider,
Ajay Shah and Lata Krishnan,
Alex and Irina Shubat,
the Silicon Valley Community Foundation,
Mary-Lou Smulders and Nicholas Hodson,
Hans Spiller,
Alan and Janet Stanford,
the Hugh Stuart Center Charitable Trust,
Clark and Sharon Winslow,
Weldon and Ruth Wood,
and many others.
KAIT and its ongoing operation were made possible by
donations from Sun Microsystems, Inc., the Hewlett-Packard Company,
AutoScope Corporation, Lick Observatory, the NSF, the University of
California, the Sylvia \& Jim Katzman Foundation, and the TABASGO
Foundation.

Support for A.V.F.'s supernova research group has been provided by the
National Science Foundation (NSF), the TABASGO Foundation, Gary and
Cynthia Bengier, the Christopher R. Redlich Fund, and the Miller
Institute for Basic Research in Science (U.C. Berkeley).
A.V.F.'s work was conducted in part at the Aspen Center for Physics, which is
supported by NSF grant PHY-1607611; he thanks the Center for its hospitality
during the supermassive black holes workshop in June and July 2018.
The U.C.S.C. group is supported in part by NSF grant AST-1518052, the Gordon \& Betty Moore Foundation,
the Heising-Simons Foundation, and by fellowships from the Alfred P.\ Sloan Foundation
and the David and Lucile Packard Foundation to R.J.F.
This research has made use of the NASA/IPAC Extragalactic Database
(NED), which is operated by the Jet Propulsion Laboratory, California
Institute of Technology, under contract with NASA.


\bibliographystyle{mnras}
\bibliography{bib}





\bsp	
\label{lastpage}
\end{document}